\documentclass[usegraphicx,useAMS,usenatbib]{mn2e}
\usepackage{times}
\usepackage{amssymb}
\usepackage{amsmath}
\usepackage{graphicx}
\usepackage{euscript}
\usepackage{rotating}
\usepackage{epsfig}
\usepackage{mathptmx}
\usepackage{amsbsy}
\def\cm{{\rm\thinspace cm}}

\def\deg{$^\circ$}
\def\erg{{\rm\thinspace erg}}

\def\ha{H$\alpha$}

\def\nii{[N{\sc ii}]$\lambda$6584}

\def\km{{\rm\thinspace km}}
\def\kpc{{\rm\thinspace kpc}}

\def\Mpc{{\rm\thinspace Mpc}}
\def\Msun{\hbox{$\rm\thinspace M_{\odot}$}}

\def\s{{\rm\thinspace s}}
\def\yr{{\rm\thinspace yr}}

\def\ergpspcmsq{\hbox{$\erg\cm^{-2}\s^{-1}\,$}}

\def\ergps{\hbox{$\erg\s^{-1}\,$}}
\def\ergpyr{\hbox{$\erg\yr^{-1}\,$}}

\def\kmps{\hbox{$\km\s^{-1}\,$}}

\def\Msunpyrpkpcsq{\hbox{$\Msun\yr^{-1}\kpc^{-2}\,$}}
\def\Msunpyr{\hbox{$\Msun\yr^{-1}\,$}}

\def\kmpspMpc{\hbox{$\kmps\Mpc^{-1}$}}

\DeclareMathAlphabet{\vib}{OML}{cmm}{m}{it}

\begin{document}
\include{defn} \title{Ionized nebulae surrounding brightest cluster galaxies}
\author[N. A. Hatch, C. S. Crawford, \& A.C.Fabian]
{N. A. Hatch$^{1,2}$\thanks{E-mail:hatch@strw.leidenuniv.nl},
C. S. Crawford$^{1}$ \& A.C.Fabian$^{1}$\\
$^{1}$Institute of Astronomy, Madingley Road, Cambridge, CB3 0HA\\
$^{2}$Leiden Observatory, P.B. 9513, Leiden 2300 RA, The Netherlands
}\maketitle
\begin{abstract}
We present integral field spectroscopic observations of six emission-line nebulae that surround the central galaxy of cool core clusters. Qualitatively similar nebulae are observed in cool core clusters even when the dynamics and possibly formation and excitation source are different. Evidence for a nearby secondary galaxy disturbing a nebula, as well as AGN and starburst driven outflows are presented as possible formation mechanisms. One nebula has a rotation velocity of the same amplitude as the underlying molecular reservoir, which implies that the excitation or formation of a nebula does not require any disturbance of the molecular reservoir within the
central galaxy.  Bulk flows and velocity shears of a few hundred \kmps\ are seen across all nebulae. The majority lack any ordered rotation, their configurations are not stable so the nebulae must be constantly reshaping, dispersing and reforming.  The dimmer nebulae are co-spatial with dust features whilst the more luminous are not.  Significant variation in the ionization state of the gas is seen in all nebulae through the non-uniform [N{\sc ii}]/\ha\ ratio. There is no correlation between the line ratio and \ha\ surface brightness, but regions with excess
blue or UV light have lower line ratios. This implies that UV from massive, young stars act in combination with an underlying heating source that produces the
observed low-ionization spectra.
\end{abstract}
\section{Introduction}
In many cool core clusters the brightest cluster galaxy (BCG) often has blue excess light indicative of recent star formation with colours that imply starbursts occurring over 0.01-1\,Gyr \citep{Allen95,Crawford1999,McNamara2004}. These galaxies are
laboratories where one can study gas cooling from the ICM and
accreting onto the galaxy, a situation which may be much
more common in the high redshift universe, and they can be sued to test theories for the growth of massive galaxies \citep[e.g][]{Benson2003,McNamara2006,Rafferty2006}.

Line-emitting nebulae surround approximately a third of all BCGs
\citep{Crawford1999}, with a strong correspondence between the
presence of an optical nebula and the short radiative cooling time of
the cluster core. Well-studied nearby examples, including NGC\,1275 in Perseus
\citep{Conselice}; NGC\,4696 in Centaurus
\citep{Crawford2005}; and A1795 \citep{Cowie1983}, exhibit extended
filamentary nebulae (up to 50\,kpc from the central galaxy), some of
which are co-spatial with soft X-ray filaments. Many of these luminous BCGs also contain reservoirs of $10^{8}-10^{11.5}$\Msun\
of molecular hydrogen 
(e.g.~\citealt{Edge2001,Salome2003}).

In this paper we present integral field spectroscopy of the ionized
nebulae that surround six BCGs. We aim to map the morphology, kinematics and
ionization state of the nebulae to gain an understanding of their
formation, heating and relationship to the cluster core. Suggested formation mechanisms include entrainment of the central molecular gas reservoir by buoyantly rising radio or ghost bubbles
\citep{BohringerM87,Churazov,Fabian2003}; the outcome of an
interaction between the gas reservoir and a secondary galaxy
(e.g.~\citealt{Bayer-Kim2002,Wilman2006}); or gas outflow induced by
a central starburst or AGN activity.  The nebulae can be extremely luminous, requiring a constant and distributed heating source \citep[e.g][]{Johnstone1988,Jaffe97}. So far this source remains unknown as no proposed heating mechanism reproduces all the
emission-line properties. A single dominant mechanism may not apply to
all BCG nebulae, and there may be a mixture of heating mechanisms
acting within a single nebula \citep [e.g][]{SabraShieldsFlip}. 

In section \ref{sample} we describe the properties of the BCG sample,
section \ref{observations} describes the observations and in section \ref{reduction} the data reduction. The morphology, kinematics and ionization state of the
nebular gas are presented in section \ref{results}, and section \ref{discussion} discusses general trends within the sample.

We have used the following set of cosmological parameters:
$\Omega_{m}=0.3,~\Omega{_\Lambda}=0.7,~{\rm H_0}=70$\kmpspMpc.
\section{Sample properties}
\label{sample}
The objects selected for the integral field unit (IFU) study were
picked from the {\it ROSAT} Brightest Cluster Sample optical follow-up
of the brightest cluster galaxies \citep{Crawford1999} and were chosen
to cover a range of optical, X-ray and radio properties. All the
target galaxies lie within the centre of clusters which exhibit bright
centrally peaked X-ray emission and have cool cores. Table
\ref{IFU_objects_prop} summaries the X-ray properties of the clusters and the radio and \ha\ luminosity of the central galaxies. We refer to the central galaxy by the name of the cluster in which it resides.

\begin{table}
 \centering 
\begin{tabular}{|l|r|r|r|r|}
  \hline
  Cluster  & $L_X$ (0.1-2.4\,keV)&1.4\,GHz power& \ha\ luminosity \\
  & $10^{44}$\ergps& $10^{24}$W\,Hz$^{-1}$&$10^{40}$\ergps\\ \hline
  A262&0.288&0.039&1.1\\
A496&1.8&0.27&11.8\\
2A~0335+096&2.25&0.104&57.2\\
RXJ\,0821+0752&2.01&0.12&$255$\\
A1068&44&4.48&$>$91.7$^{1}$\\
A2390&13.4&10&120\\
\hline
\end{tabular}

\caption{\label{IFU_objects_prop}Properties of clusters and galaxies
in IFU study.  References for X-ray luminosity:  \citet{Ebeling1996}; \citet{Ebeling1998}.  References for radio power: \citet{Birzan2004}; \citet{Markovic2004}, \citet{Sarazin1995}, \citet{Becker1995}; \citet{Gubanov1999}; \citet{Ebeling1998}. References for \ha\ luminosity: \citet{Heckman}; \citet{Cowie1983}; \citet{Romanishin}; \citet{Bayer-Kim2002}; \citet{Allen1992}. $ ^{1}$ luminosity is given as lower limit as it has been
determined from slit spectroscopy}

\end{table}
\begin{description}
\item[{\bf A262}] is a relatively poorly populated nearby cluster with a mass deposition ratio of 19\Msunpyr \citep{Blanton2004}. The central radio
source B2\,0149+35 has a double-lobed morphology orientated
approximately East-west \citep{Parma1986}. The interaction between the radio source and the ICM has created two cavities seen as low surface brightness
regions in X-ray images that coincide with the radio lobes
\citep{Blanton2004}.  With a redshift of z=0.0163, the linear scale gives
0.33\,kpc\,arcsec$^{-1}$.
\item[{\bf A496}] is a relaxed cluster with a cool core with a central
metal abundance enhancement \citep{Tamura2001}. The central cD galaxy
(MCG --02-12-039) is a fairly weak line-emitter \citep{Fabian1981,Hu}
and is host to the compact radio source MSH 04-112
\citep{Markovic2004}. At a redshift of 0.0329 the galaxy has a
linear scale of 0.656\,kpc\,arcsec$^{-1}$.
\item[{\bf 2A~0335+096}] is an X-ray bright cluster in which the cluster
core shows complex structure including depressions in the X-ray
surface brightness, interpreted as `ghost bubbles' \citep{Kawano2003}
and soft X-ray filaments that align with the emission-line nebula
\citep{Sarazin1992}. A weak elongated radio source is associated with the
central galaxy, with two emerging jets at position angles 60$^{\circ}$
and 240$^{\circ}$ in the direction of the X-ray cavities \citep{Sarazin1995}.  \citet{Romanishin} were the first to image the filamentary nebula, showing a 12\,kpc extension in the direction of a nearby secondary galaxy that lies 6.5\,arcsec (4.5\,kpc) to the
Northwest of the BCG. At a redshift of 0.0349,
1\,arcsec corresponds to 0.695\,kpc.
\item[{\bf RXJ\,0821+0752}] The central galaxy is host to a weak radio source, is surrounded by bright blue knots to the Northeast of the galaxy and is known to be a luminous \ha+[N{\sc ii}] emitter \citep{Bayer-Kim2002}. The galaxy contains a large central molecular gas reservoir \citep{Edge2003}.  The X-ray surface brightness is centred on the BCG and extends  to the Northwest in the same direction as the \ha\ and CO emission. Assuming a steady-state cooling flow the cluster has a mass deposition rate of $\sim30$\Msunpyr. \citet{Bayer-Kim2002} suggest that the emission-line nebula may be formed by a combination of processes including ICM cooling and disturbance of the gas reservoir by the nearby secondary galaxy that lies 4\,arcsec to the Southeast of the central galaxy. The galaxy has a redshift of 0.11 \citep{Crawford1999}.
\item[{\bf A1068}] A rich cluster in which the cD galaxy is offset
from the X-ray centroid by approximately 2\,arcsec
\citep{McNamara2004}.  This high X-ray luminosity cluster exhibits complex ICM structure in the core. The {\it Chandra} data are consistent with a cooling rate of 30\Msunpyr within a core radius of 30\,kpc \citep{Wise2004}.  The galaxy colours show that the BCG has been experiencing star formation at the rate of 20-70\Msunpyr for the past
100\,Myr and 95\% of the ultraviolet and \ha\ photons emerge from the
region with the shortest X-ray cooling time, possibly indicating that
the ICM is fueling the star formation in the central galaxy
\citep{McNamara2004}.  The nebula emits strong [O{\sc iii}] line
emission indicating that the source of ionization is different to
other BCGs which have weak [O{\sc iii}] emission beyond the galaxy nucleus.  Weak
Wolf-Rayet features in the spectrum suggest that the dominant source
of nebular ionization is a massive starburst \citep{Allen95}. 
A1068 has a redshift of 0.1386 giving a linear scale of 2.45\,kpc\,arcsec$^{-1}$.
\item[{\bf A2390}]
This rich cluster has a high X-ray luminosity and a large mass deposition rate of
200-300\,\Msunpyr \citep{Allen2001}. The brightest galaxy lies
at the peak of the X-ray emission, and has extended optical
line-emission \citep{LeBorgne1991}, Lyman-$\alpha$ emission
\citep{Hutchings2000} and contains dust \citep{Edge1999}.  The galaxy
is also host to a core-dominated radio source \citep{Augusto2006}
which is one of the most luminous radio sources within a BCG. The radio source is a
medium-sized symmetric object orientated North-South, misaligned
approximately 45 degrees from the emission-line structure.  At 1.65GHz
the radio emission has an extent of 0.9\,arcsec \citep{Augusto2006},
although the resolution of the radio observations is such that any
structure on scales of $\sim$10 arcsec could be missed. The galaxy has
a visible star formation rate of 5.4\Msunpyr \citep{Crawford1999}.
A2390 is the most distant cluster in the sample, with a redshift of
z=0.2301, the luminosity distance is 1147\,Mpc and 1\,arcsec
corresponds to 3.6\,kpc.\\
\end{description}

\section{Observations}
\label{observations}
The data were obtained with the Oasis integral field unit (IFU) on the William
Herschel Telescope (WHT) on the nights of 2005 Sept 17, 22, 27, Dec 6, 2006 Dec 1,2. The seeing conditions were 0.8-1.5\,arcsec. The NAOMI adaptive optics system was used throughout the observations
to improve the strehl ratio of the observations. Significant
improvement can be made if a bright guide star is situated nearby the
target galaxy, such as near A262. No bright (${\rm m}_{V}>12$) guide
star was available for the other five targets, therefore only fast
tip-tilt correction was used with the available fainter guide stars. In addition to the science targets the standard star HR7950 was
observed with the MR807 grating and star HR1544 was observed with the
MR661 grating on the night of Sept 17. The seeing at the time of the A262, 2A\,0335+096, and A2390 observations was measured to be 0.7\,arcsec from the reconstructed image of the standard star taken with the OASIS IFU. The \ha\ and continuum emission is almost point-like in the nucleus of A1068 therefore a Gaussian model was fit to the \ha\ nucleus of the galaxy and the FWHM of the seeing is measure to be 1.1\,arcsec. The seeing during the A496 and RXJ\,0821-0752 were measure by the DIMM to be $<1.4$\,arcsec.  The spatial configuration was determined by using the 22\,mm enlarger, which gave an IFU field-of-view of $10.3\times7.4$\,arcsec. This area
was divided into $\sim$1100 lenslets resulting in a spatial sampling
of $0.26\times0.26$\,arcsec. Each target galaxy was observed in 2 or more
exposures, each of approximately 900\,sec. Each exposure was offset by 0.4\,arcsec to
provide oversampling and avoid bad pixels on the CCD. The airmass of A262, A2390 and 2A~0335+096 was $\le$1.06 throughout the observations, whilst A1068 was observed at an airmass $\le$1.13 and A496 at an airmass of $\le$1.35. Due to these small airmasses, observing in the red part of the spectrum and over a very narrow
wavelength range, there was no need to correct for the relative shift
due to the atmospheric dispersion \citep{Filippenko1982}. Details of the observations are summarized in Table \ref{tab:observations}. The spectral resolution ranges between $223-273$\kmps. The MR\_661 grating observes the wavelength range 6210--7010\AA, MR\_735 observes 6940--7760\AA, whilst MR\_807 observes the range 7690--8460\AA.

\begin{table*}{
\centering

\begin{tabular}{|l|l|l|l|l|l|l|}
\hline
Cluster  &RA& Dec&Redshift&Exposure  &Grating&Seeing\\
name &(J2000)&(J2000)&&time (sec)&(arcsec)&\\ \hline
A262&01 52 46.5& 36 09 08&0.0163&4$\times$900&MR\_661&0.7\\

A496&04 33 37.7&-13 15 39&0.0329& 1000+900 &MR\_661& -- \\ 

2A0335+096&03 38 40.5&09 58 12&0.0349&4$\times$900&MR\_661&0.7\\

RXJ0821+0752&08 21 02.6& +07 51 31&0.11&6$\times$900&MR\_735&$<$1.3\\

A1068&10 40 44.4&39 57 12&0.1375&6$\times$900&MR\_735&1.1\\

A2390&21 53 36.7&17 41 45&0.228&4$\times$900&MR\_807&0.7\\
\hline
\end{tabular}
\caption{Summary of observations.  The seeing is the FWHM
and is given in arcsec. The error on the seeing is $\sim20$ per
cent.\label{tab:observations}}}
\end{table*}
 \section{Data reduction}
 \label{reduction}
The data were processed using the OASIS dedicated reduction package
XOASIS (version 6.3).  Each datacube underwent basic
data-reduction steps including overscan correction, bias subtraction, spectra extraction,
wavelength calibration, flat-fielding, cosmic ray removal and
sky-subtraction using either an average sky spectrum extracted from an
area of empty sky or using a 900 second blank sky frame.  The precision of the wavelength calibration was checked. 13 arc lines were detected in the
MR\_661 grating. The mean error was 0.017\AA\ with a standard deviation
of 0.01\AA. For the MR\_807 grating only 5 arc lines were detected and
the mean error was 0.005\AA\ with a standard deviation of
0.003\AA. Seven arc lines were detected in the wavelength range of the
MR\_735 grating, the mean error was 0.061\AA\ with a standard
deviation of 0.028\AA. The data cubes were shifted to remove
the 0.4\,arcsec dither, re-normalised to account for variations in transparency, resampled to a spatial scale of 0.2$\times$0.2\,arcsec per lenslet, and finally median combined
(except A496 which was taken as the mean of the two
observations). The datacubes of A262, A2390 and 2A~0335+096 were flux calibrated using the standard stars HR1544 and HR7950.  

The data of each lenslet were fitted with a multiple Gaussian and constant continuum model for the lines of \ha, [N{\sc ii}]$\lambda\lambda$6548, 6584, [O{\sc
i}]$\lambda\lambda$6300, 6363, and [S{\sc ii}]$\lambda\lambda$6717,
6731. All the emission-lines
were forced to have the same redshift and velocity width. The
continuum level was obtained by fitted a constant over
the continuum region between the [O{\sc i}]$\lambda$6363 line and the
[N{\sc ii}]$\lambda$6548 line. The continuum parameter was then frozen
which acted as to remove the continuum level. The flux of
the [O{\sc i}]$\lambda$6363 line was set at one third of the flux of
the [O{\sc i}]$\lambda$6300 line, and the relative normalisation of
[N{\sc ii}]$\lambda$6548 was fixed at a third of the [N{\sc
ii}]$\lambda$6584 emission-line.

Data from A2390 did not cover the [S{\sc ii}] and data from
2A~0335+096 did not cover the [O{\sc i}] lines therefore the Gaussian
normalisation of these lines were fixed at zero during the line
fitting process. Line fitting was done using the {\sc qdp} package
\citep{qdp}. 

Maps of the \ha+[N{\sc ii}] flux, line-of-sight velocity, line-width, and
[N{\sc ii}]/\ha\ line ratio were created for each nebula. The line-of-sight velocities of
the emission line nebulae were obtained from the Doppler shifts of the
strong \ha\ and [N{\sc ii}] emission-lines. The zero-point of the
line-of-sight velocity is defined as the redshift of the 
lenslet at the galaxy centre. A cross is placed on each image identifing the IFU
lenslet at the galaxy centre. The galaxy centre is defined as the lenslet
containing the maximum continuum emission, except in A262.  A large
dust lane cuts North-South across A262, therefore the continuum emission from the central region is fainter than the surrounding area.  Instead the IFU lenslet with
the maximum \ha+[N{\sc ii}] emission marks the galaxy centre. This
central lenslet is situated in the line-width peak, which corroborates
its identity as the galaxy centre. All line-widths discussed are FWHM (full width at half
maximum) and have been corrected for instrumental broadening. The FWHM
of the instrumental profile is $\sim$215-260\kmps at \ha\ which was
determined from nearby sky-lines and arc lamp line emission. In these maps we present only the pixels in which the flux in the strongest lines (\ha\ and [N{\sc ii}]) were greater than three times the uncertainty (from the goodness-of-fit between the Gaussian model and the data) of the  intensity parameter of each line. This 3$\sigma$ criterion limits the spatial extent of the IFU maps. For A496 only [N{\sc ii}]$\lambda$6584 is particularly strong, therefore only this line is used to determine whether line-emission is present. 
The detection limit is approximately $10^{-16}$\ergpspcmsq, so there may be a fainter component that is not visible in these images.

\section{Results}
\label{results}
\subsection{A262}
\label{A262_results}
\begin{figure*} \centering
\includegraphics[width=0.3857\columnwidth]{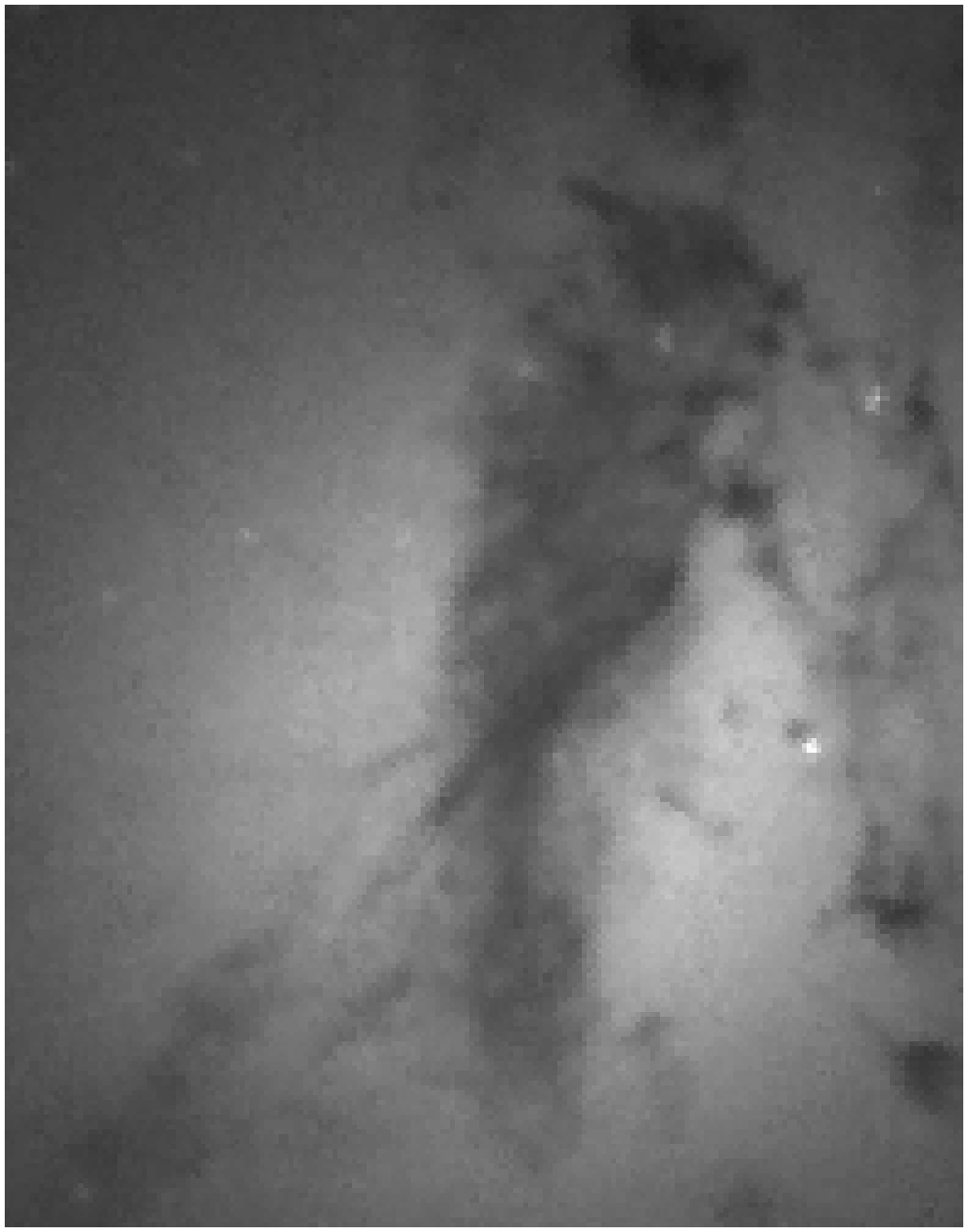}
\includegraphics[width=0.4\columnwidth]{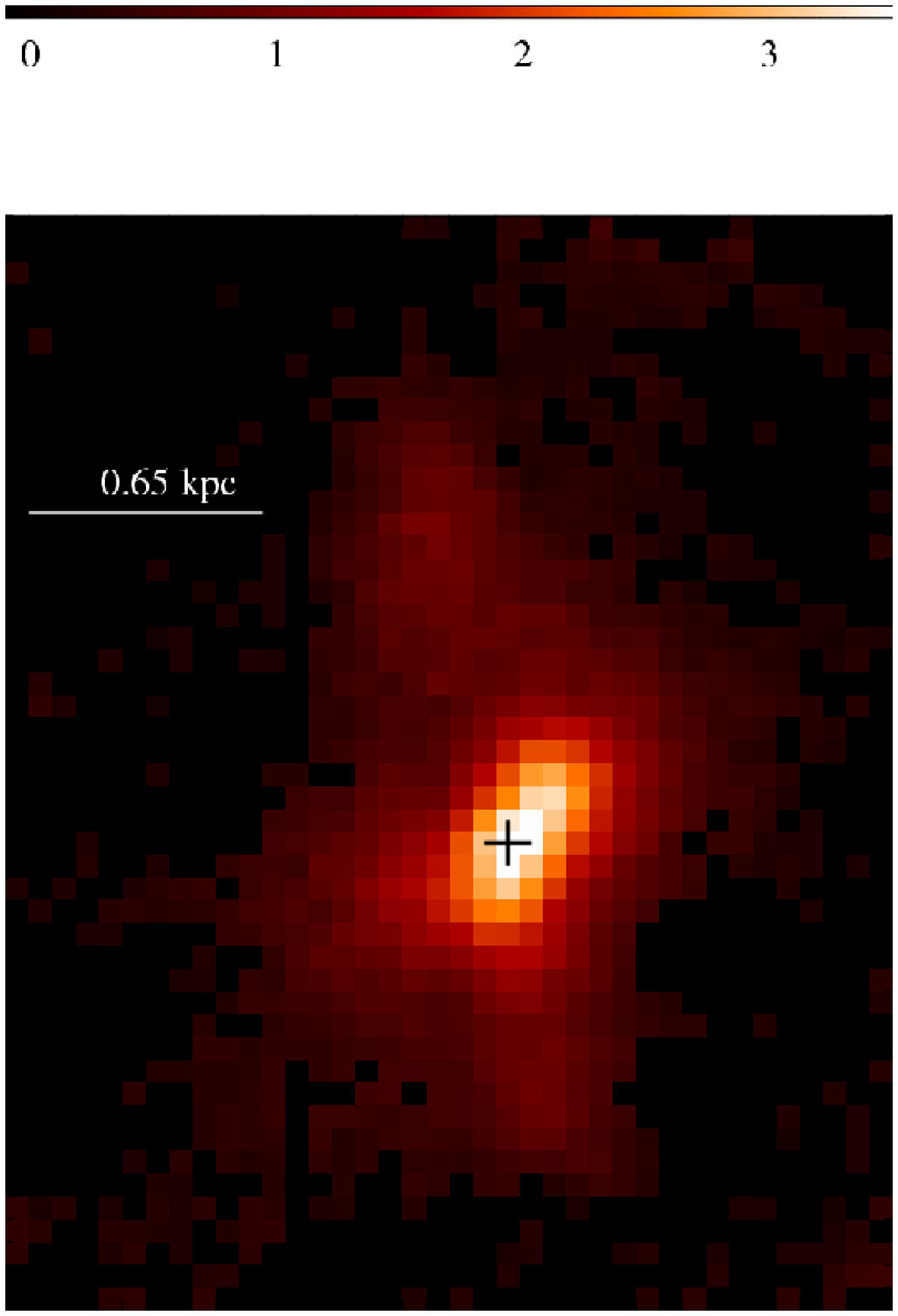}
\includegraphics[width=0.4\columnwidth]{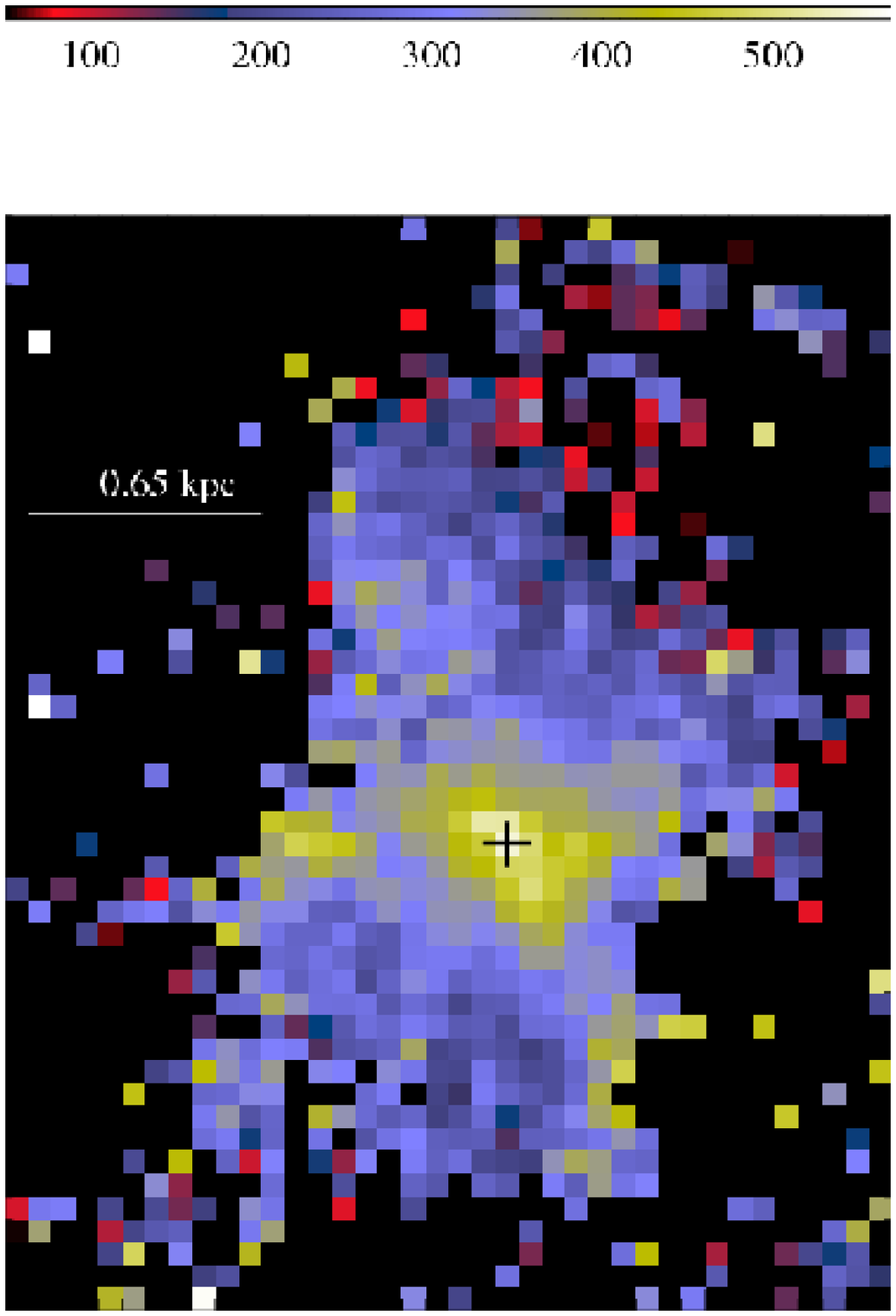}
\includegraphics[width=0.4\columnwidth]{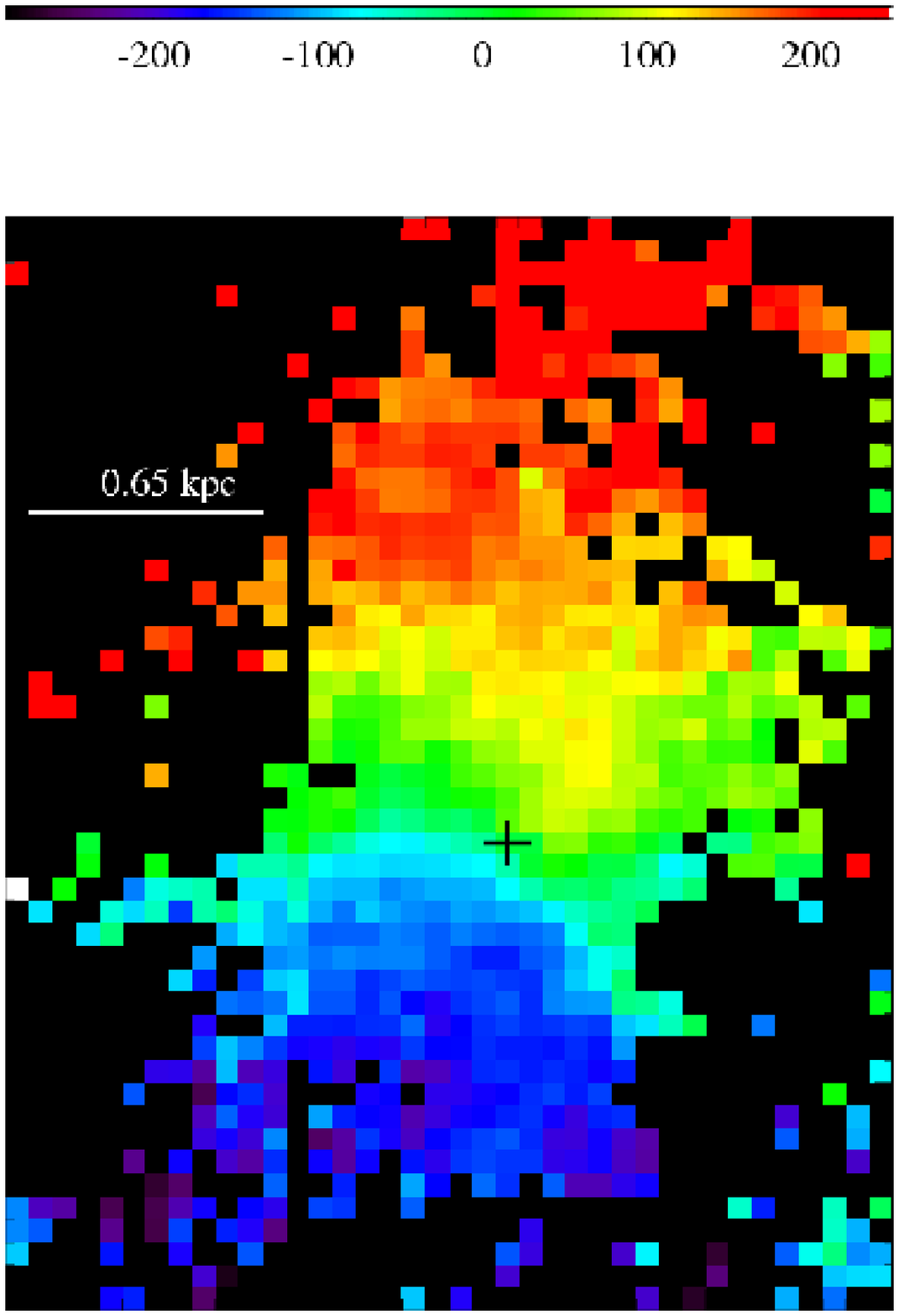}
\includegraphics[width=0.4\columnwidth]{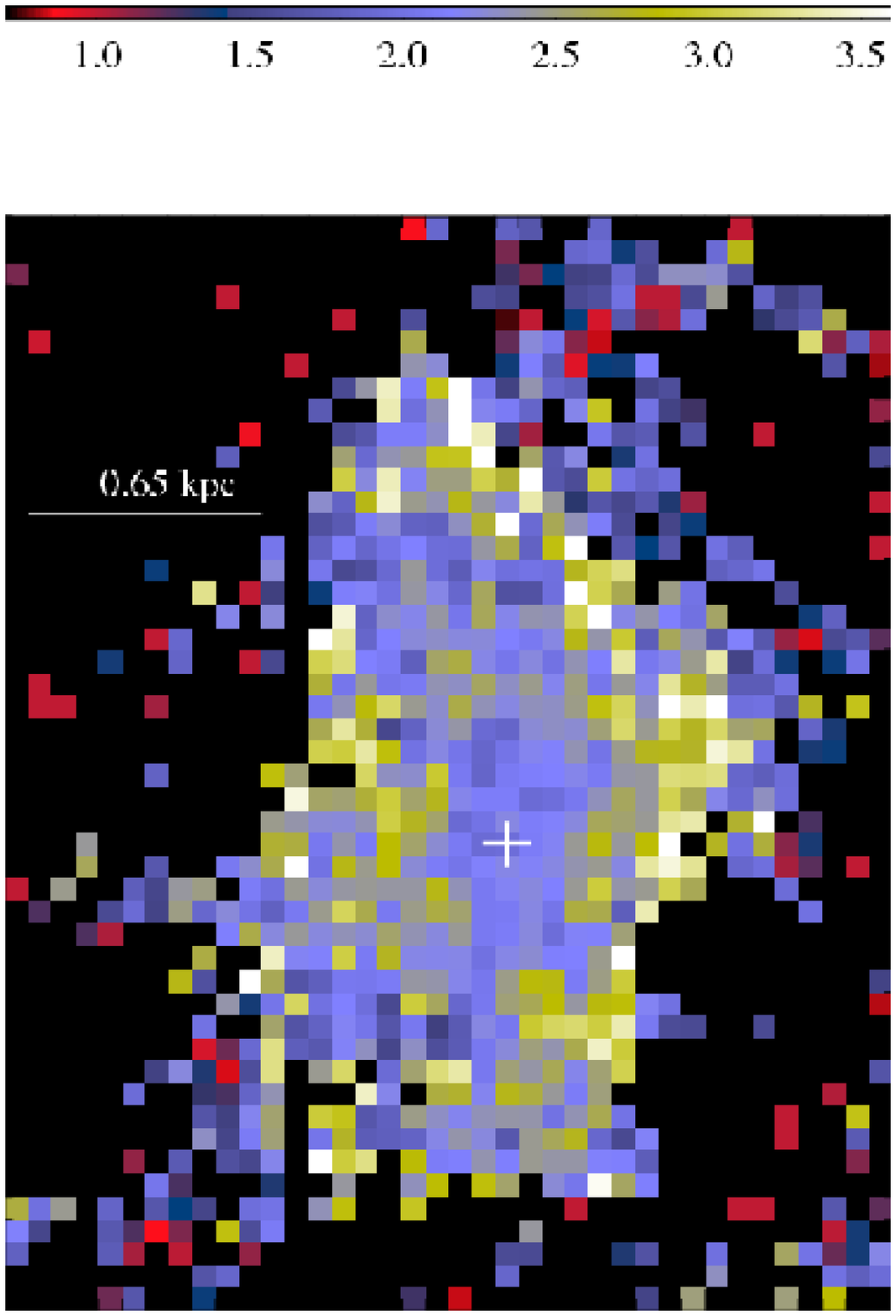}
\caption{Images of A262 from left to right: HST ASC F435w; \ha +[N{\sc ii}] flux (with colour bar scaling in units of 10$^{-16}$\ergpspcmsq); line-width
  [\kmps]; line-of-sight velocity [\kmps]; [N{\sc ii}]/\ha\ ratio. Images are 7.4$\times$9.4\,arcsec. In all Figs.~\ref{A262_ifu}--\ref{A2390_ifu} (except \ref{A496_ifu}) only the IFU lenslets in which \ha\ and [N{\sc ii}]$\lambda$6584 are detected  above 3$\sigma$ are presented. The cross marks the emission-line flux peak which indicates the galaxy centre. North is up, East is left in all Figs.~\ref{A262_ifu}--\ref{A2390_ifu}.
  \label{A262_ifu}}
\includegraphics[width=0.41\columnwidth]{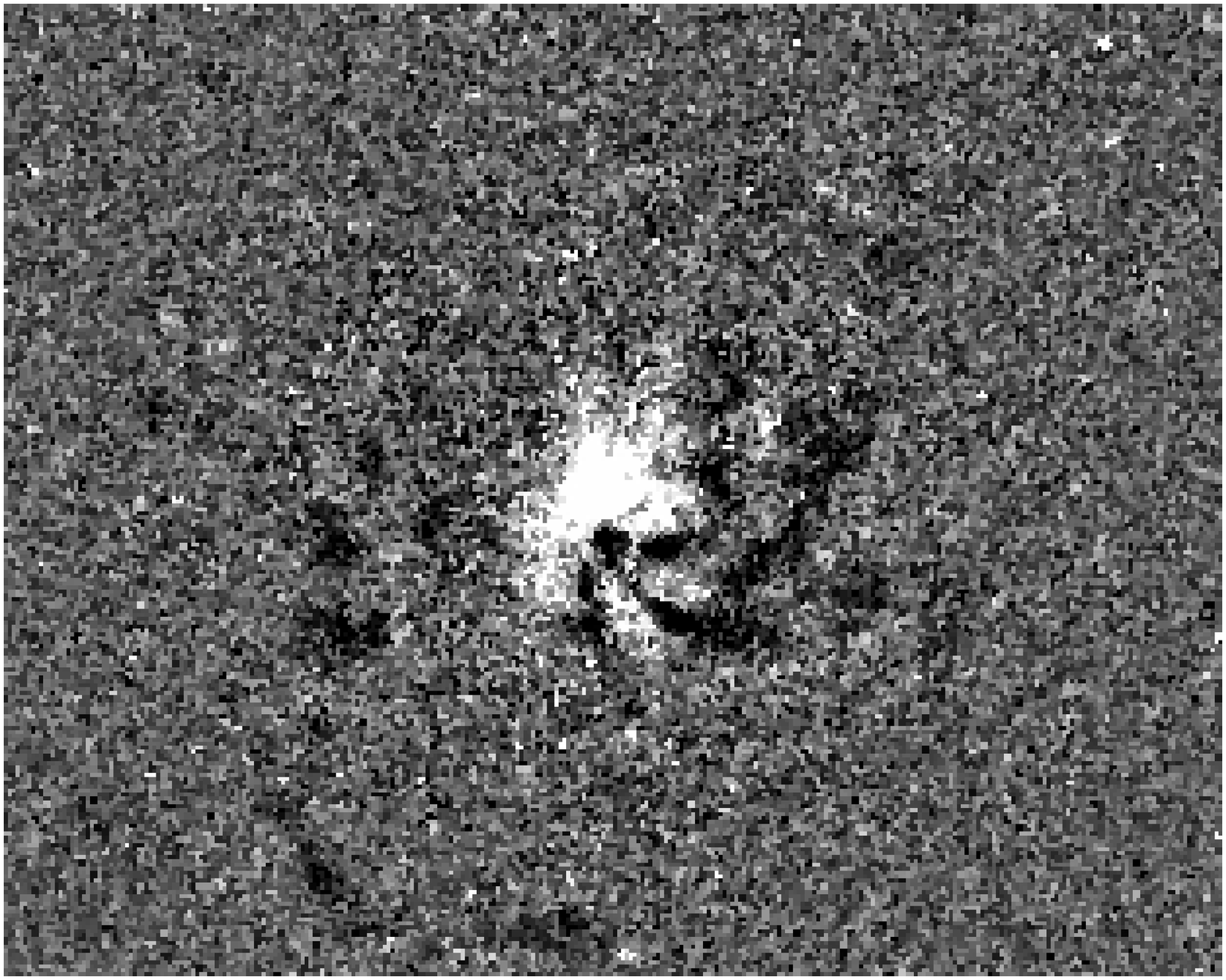}
\includegraphics[width=0.4\columnwidth]{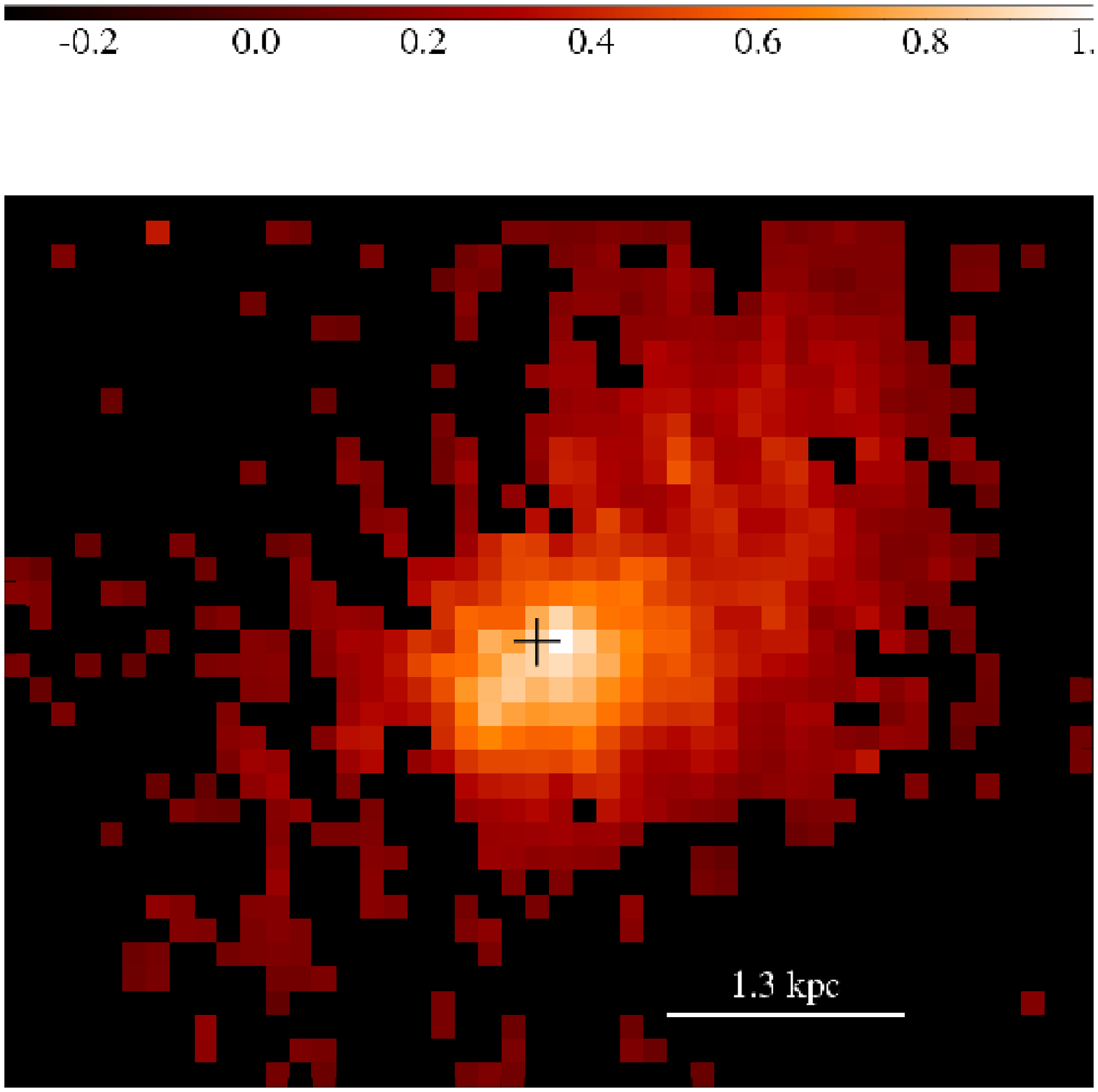} 
\includegraphics[width=0.4\columnwidth]{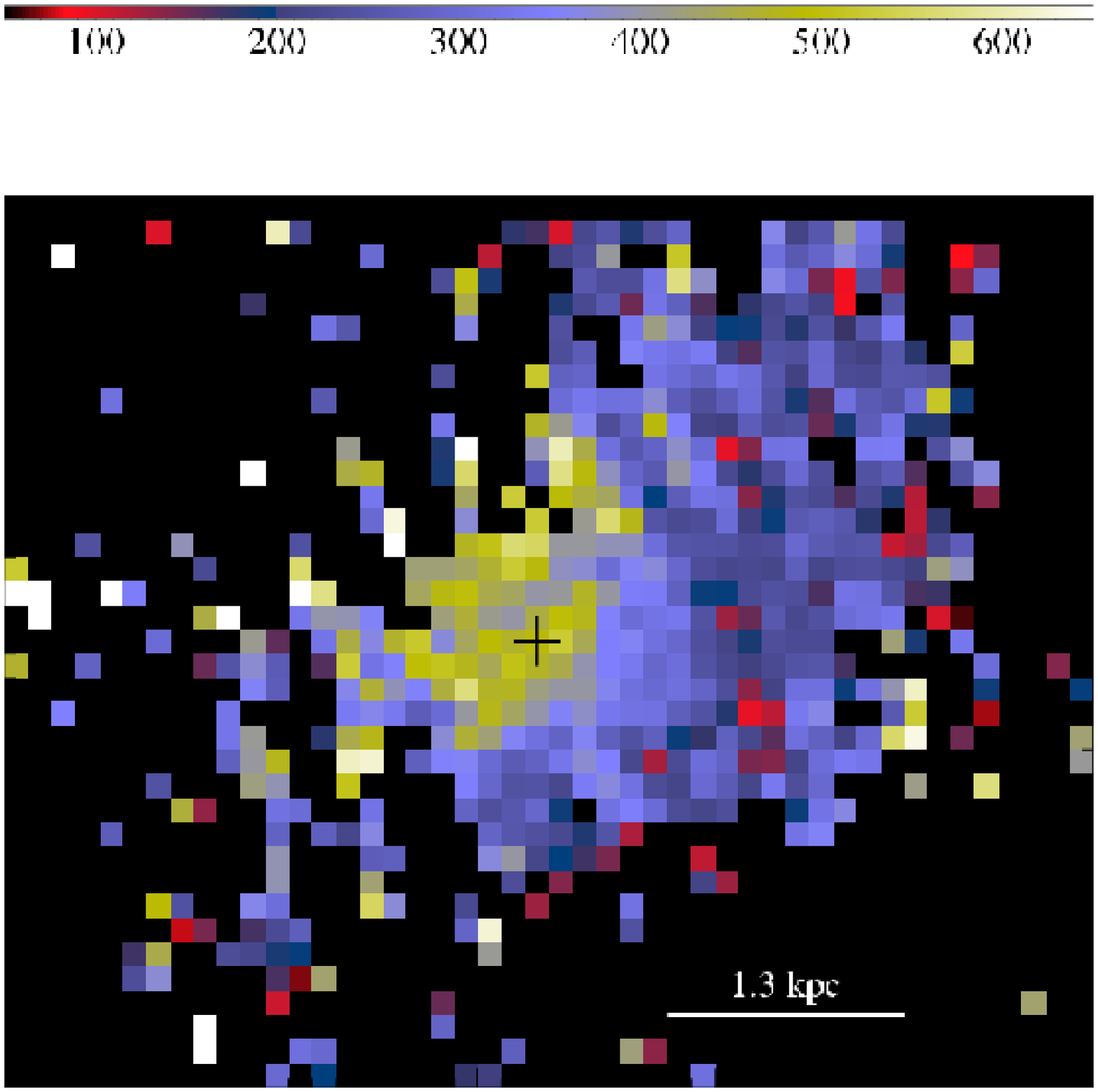}
\includegraphics[width=0.4\columnwidth]{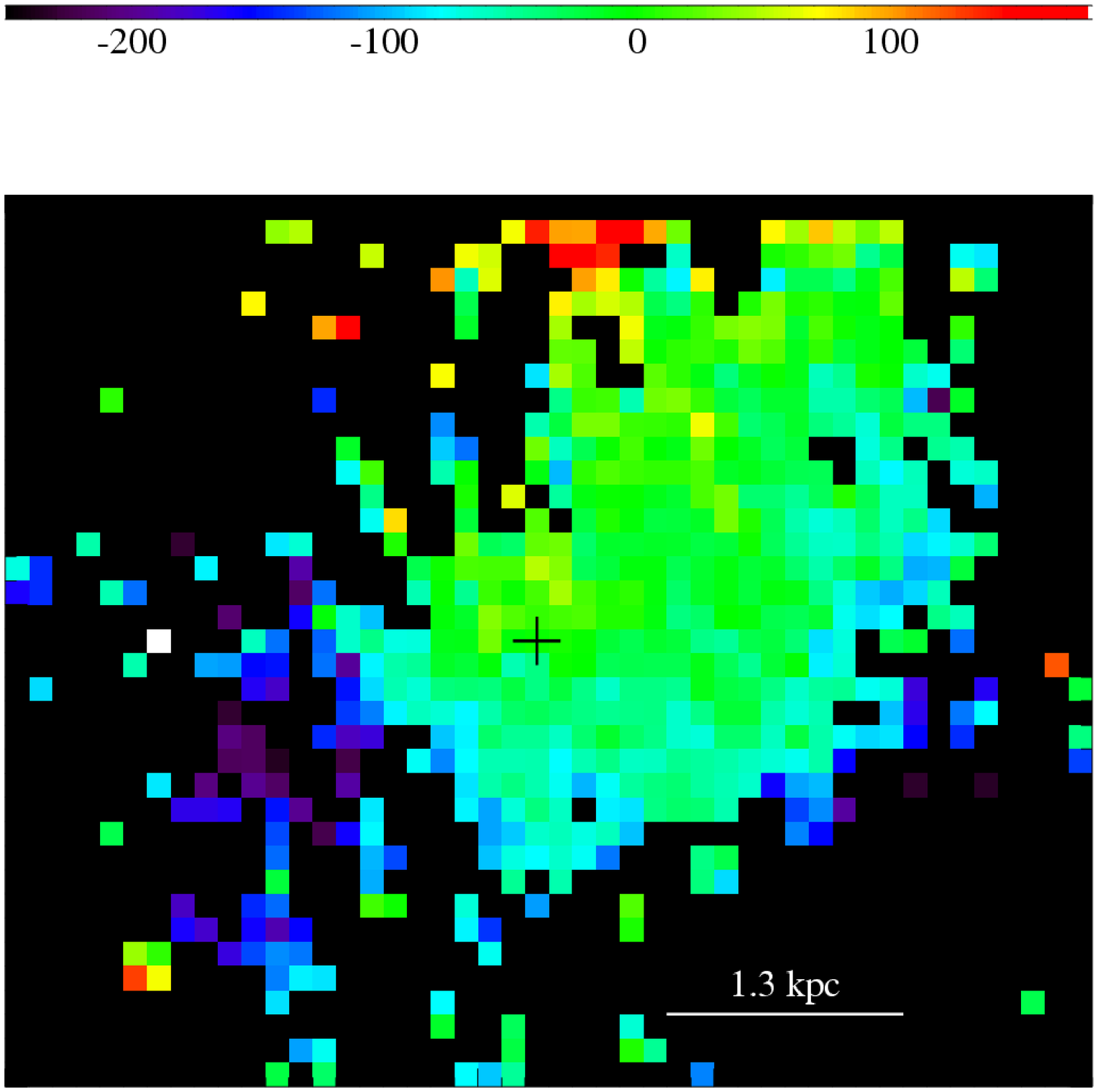}
\includegraphics[width=0.4\columnwidth]{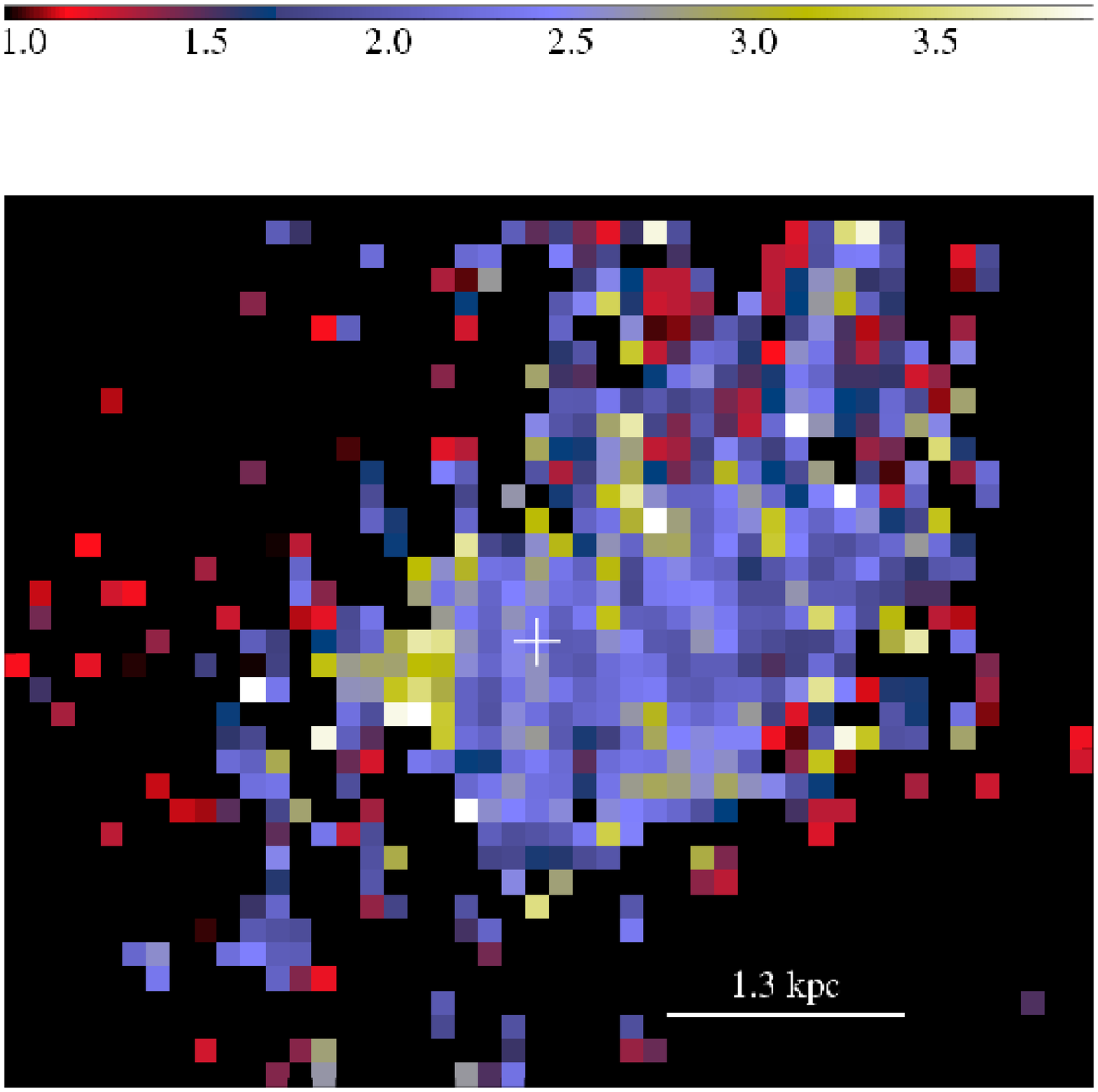}
\caption{Images of A496 from left to right: Unsharp mask HST ASC F702w image; normalized \ha +[N{\sc ii}] flux; line-width [\kmps]; line-of-sight velocity [\kmps]; [N{\sc ii}]/\ha\ line ratio. Images are 9.0$\times$7.4\,arcsec. Only lenslets in which [N{\sc ii}]$\lambda$6584 is detected above 3$\sigma$ are presented.  The cross marks the continuum maximum which indicates the galaxy centre
\label{A496_ifu}}
  \includegraphics[width=0.50\columnwidth]{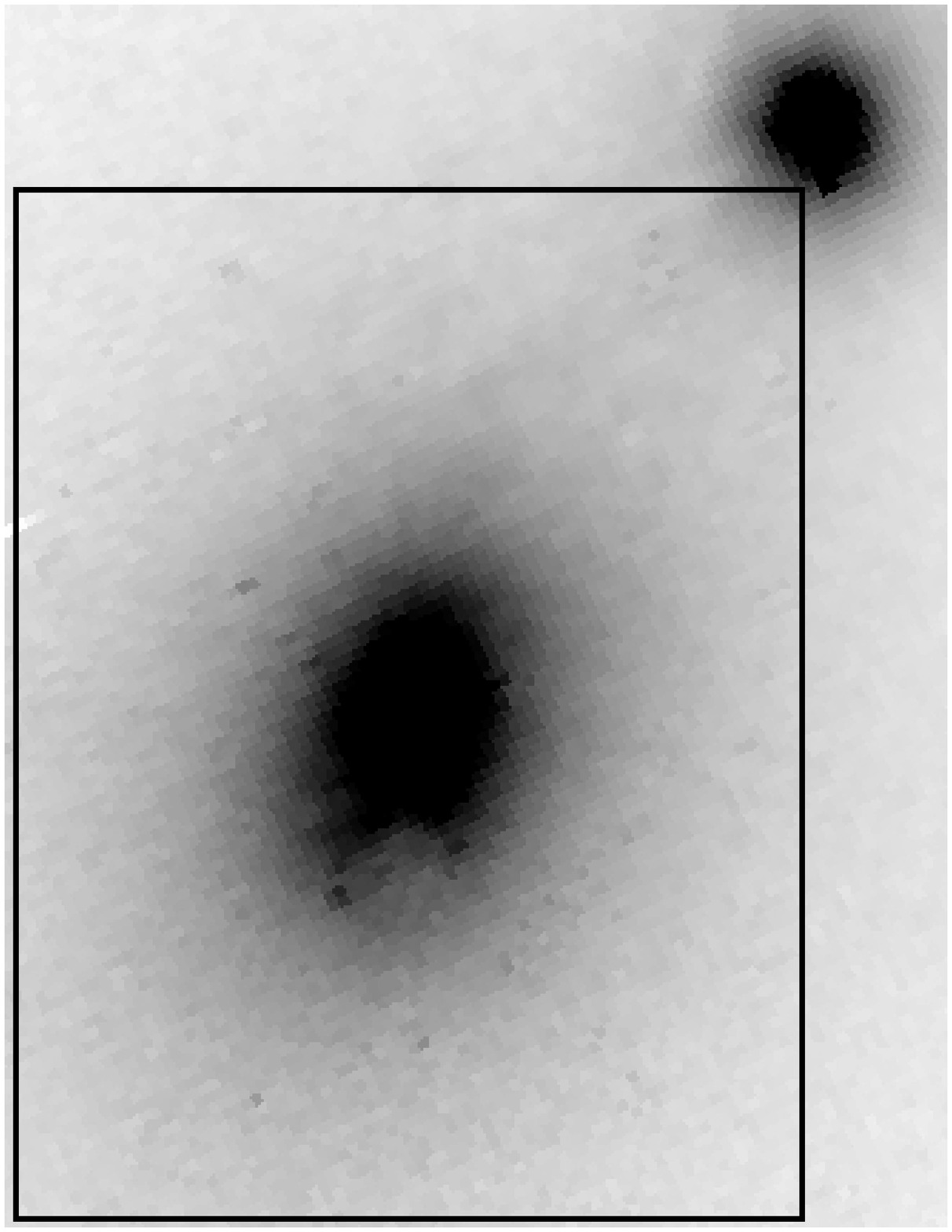}
  \includegraphics[width=0.38\columnwidth]{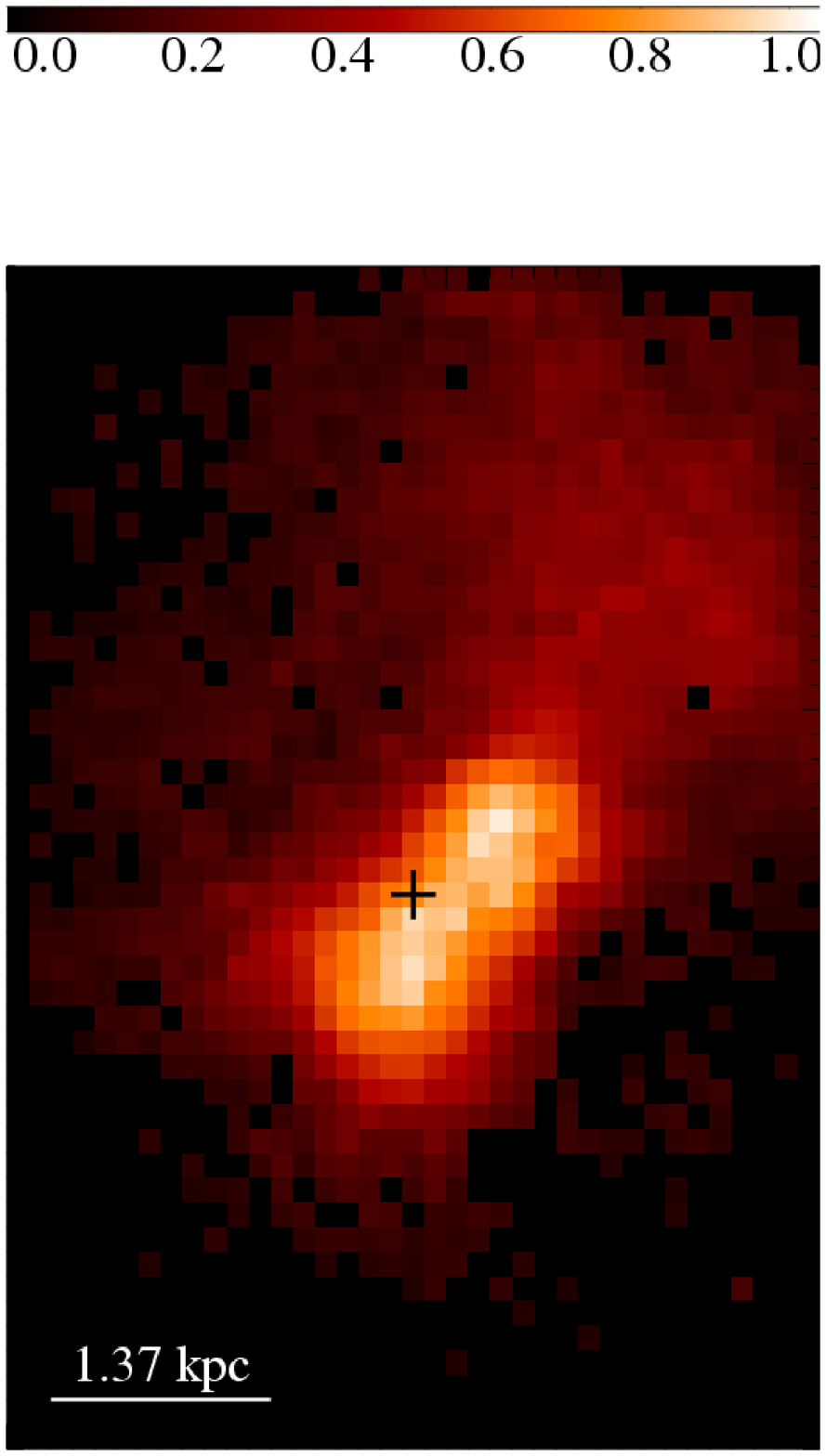}
  \includegraphics[width=0.38\columnwidth]{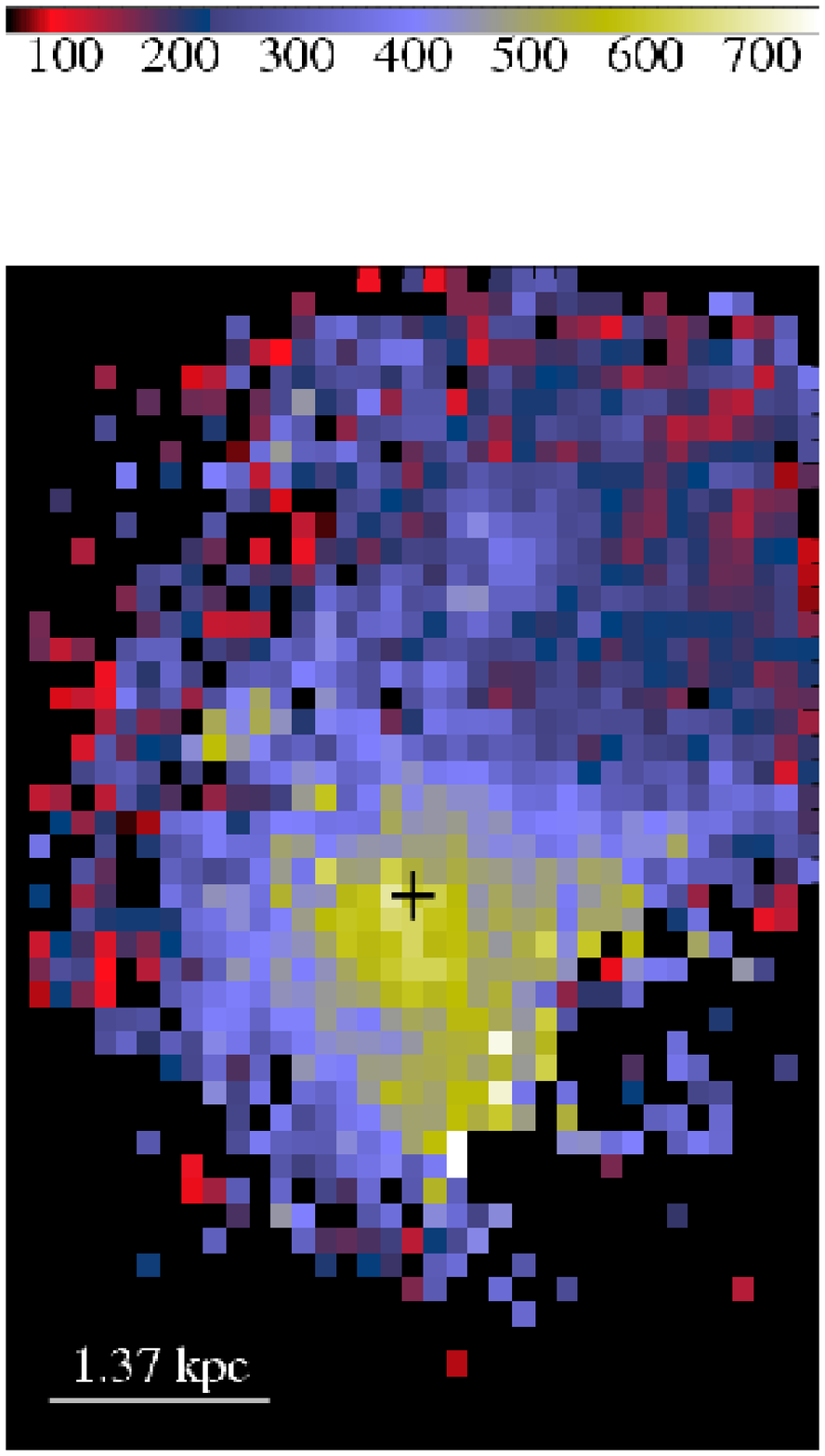}
  \includegraphics[width=0.38\columnwidth]{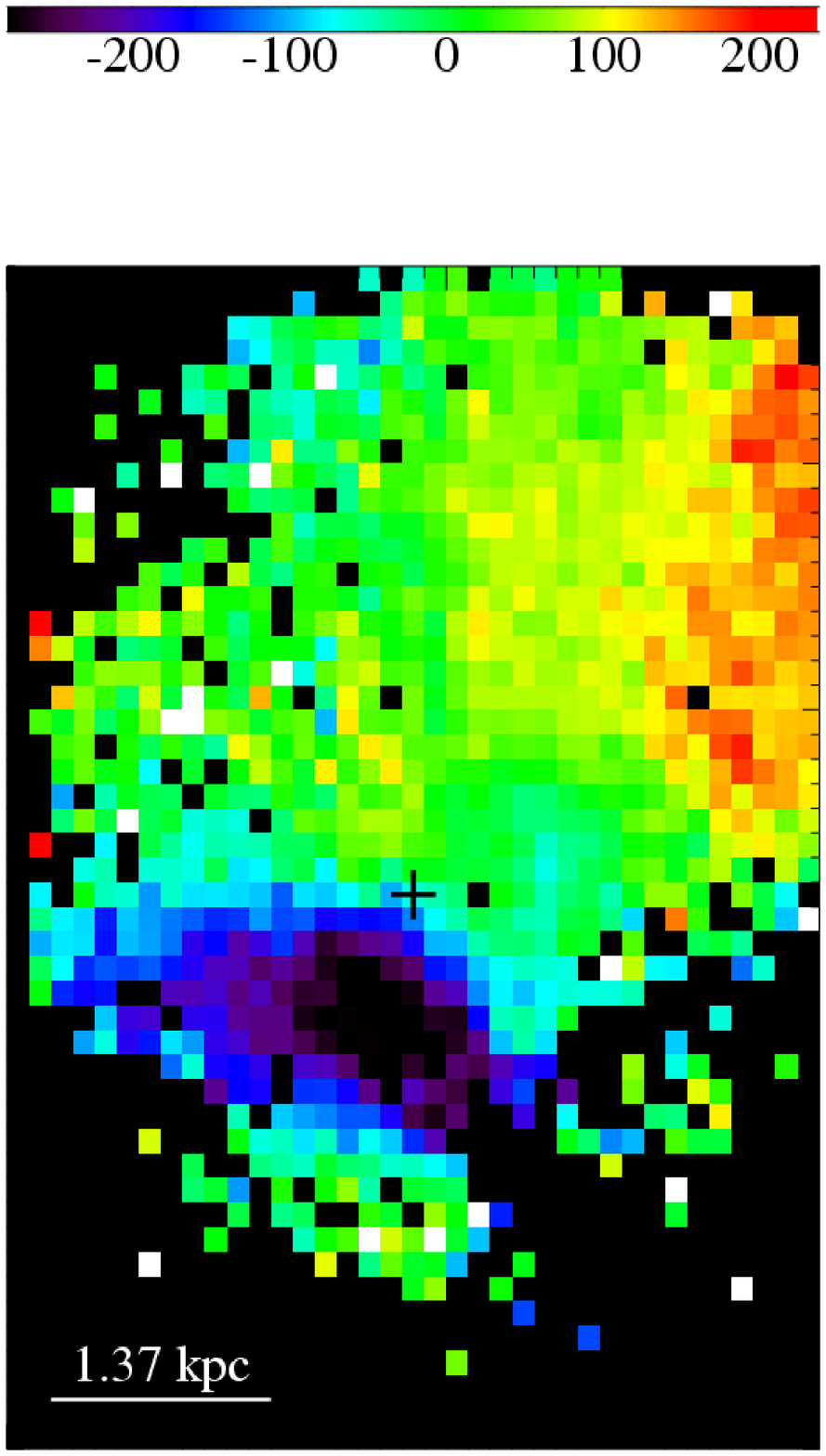}
 \includegraphics[width=0.38\columnwidth]{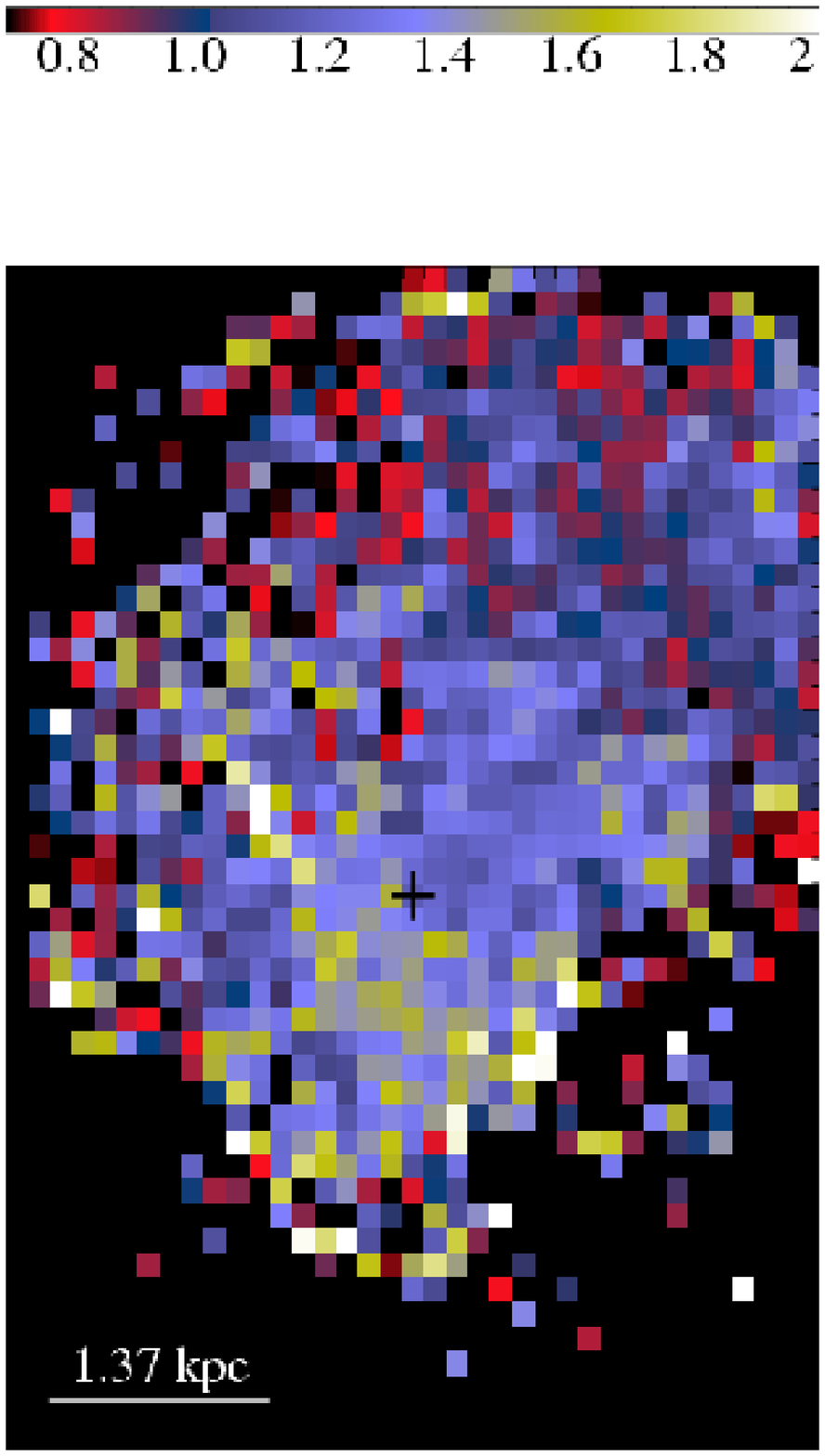}
 \caption{Images of 2A~0335+096 from left to right: negative HST ASC F606w; \ha +[N{\sc ii}] flux (with colour bar scaling in units of 10$^{-16}$\ergpspcmsq); line-width [\kmps]; line-of-sight velocity [\kmps]; [N{\sc ii}]/\ha\ ratio. Images are 7.2$\times$9.4\,arcsec.  The black box indicates the IFU field-of-view. A secondary galaxy is visible to the Northwest just beyond the IFU field-of-view. The cross marks the galaxy centre defined by the continuum maximum.   \label{RXJ0335_ifu}}
\end{figure*}
\begin{figure*}
\centering
    \includegraphics[width=0.415\columnwidth]{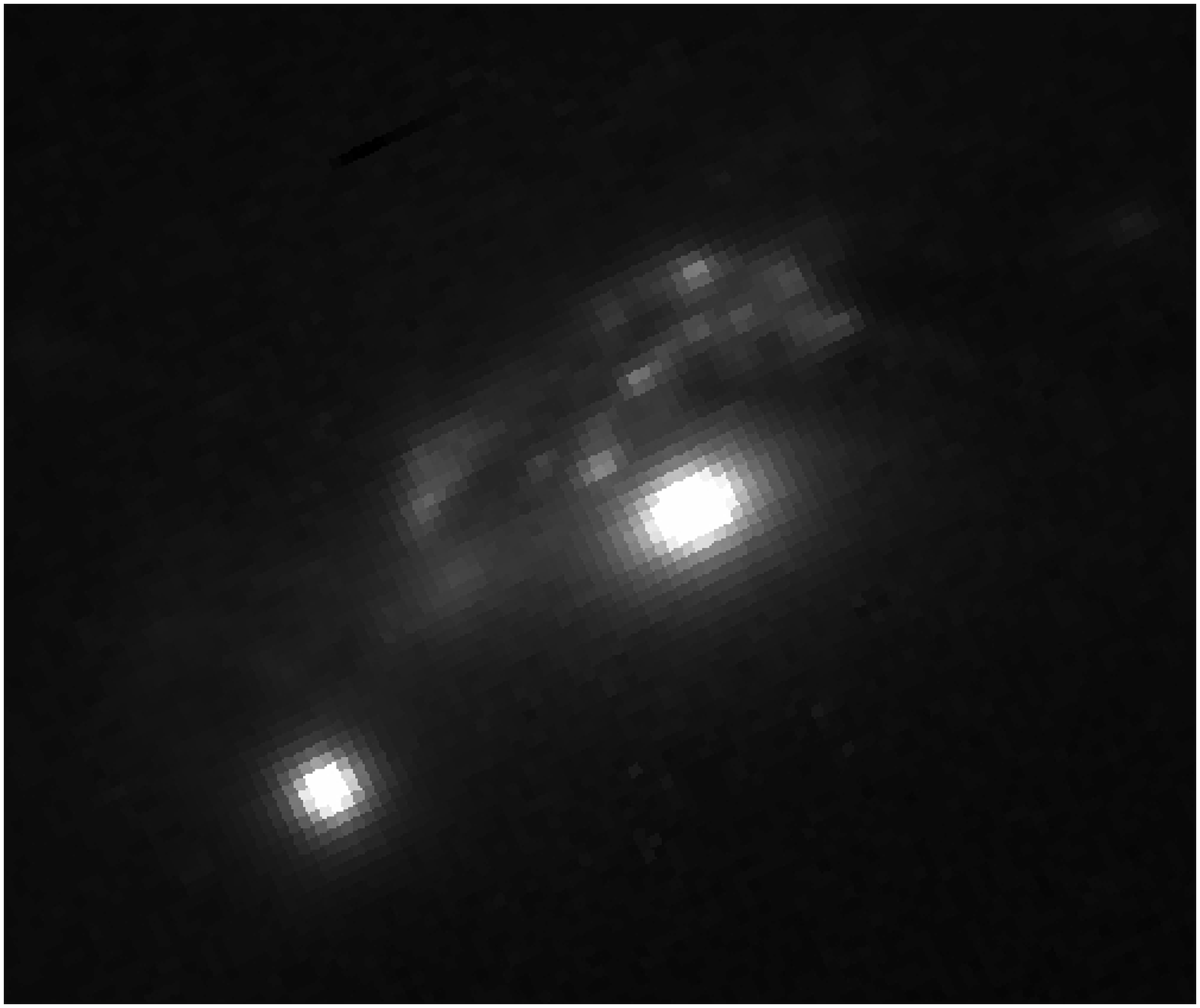}
    \includegraphics[width=0.4\columnwidth]{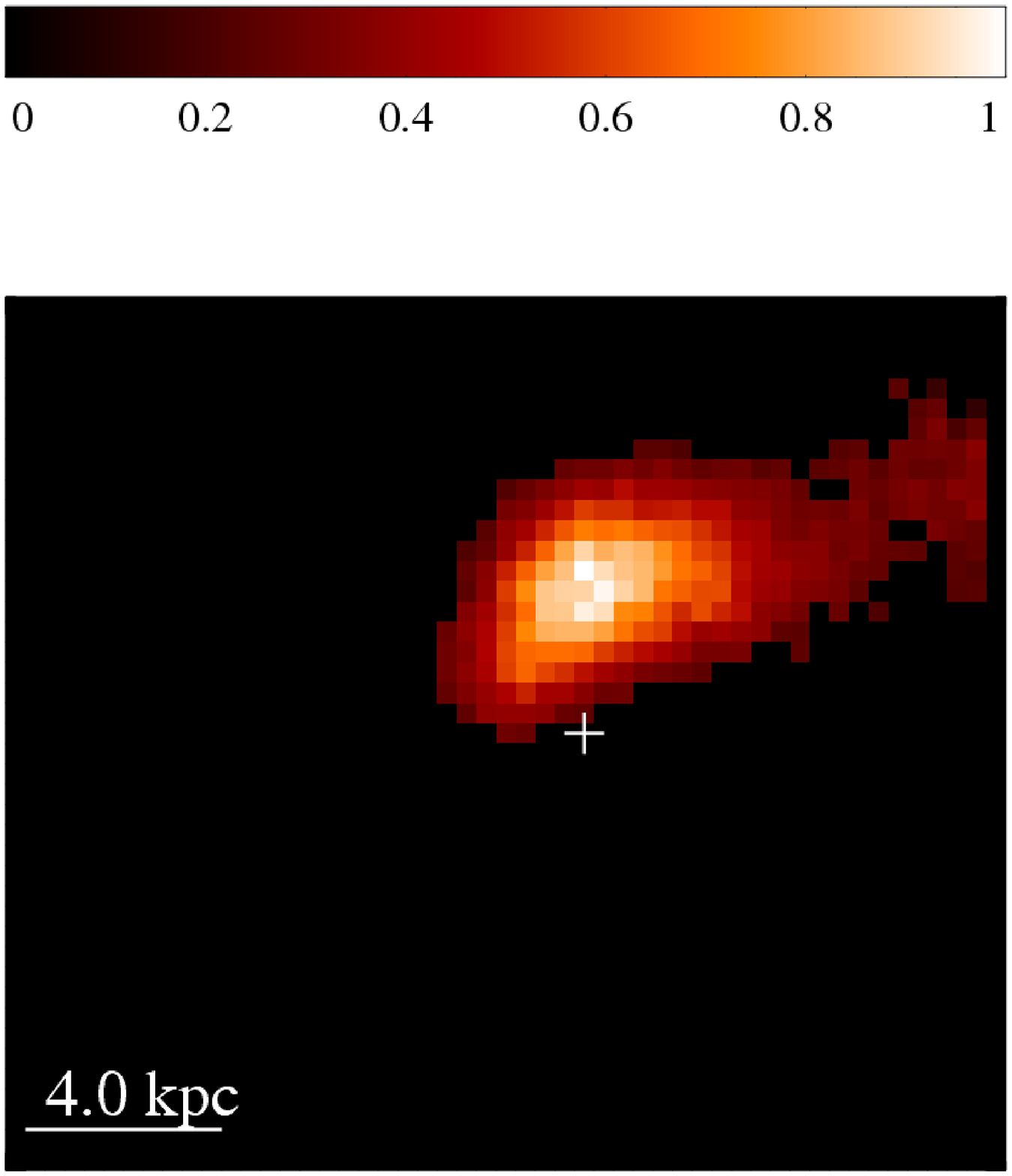}
  \includegraphics[width=0.4\columnwidth]{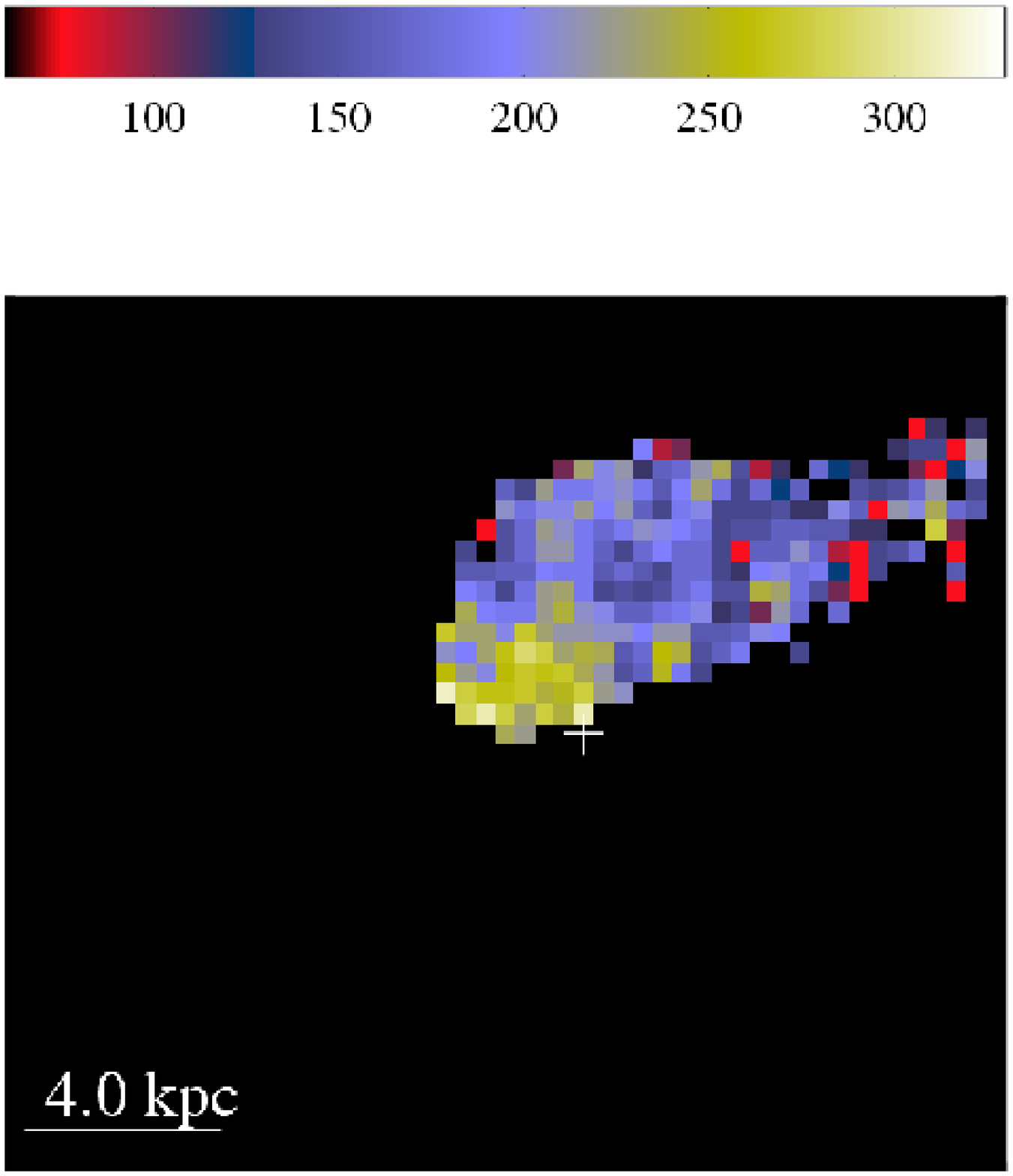}
  \includegraphics[width=0.4\columnwidth]{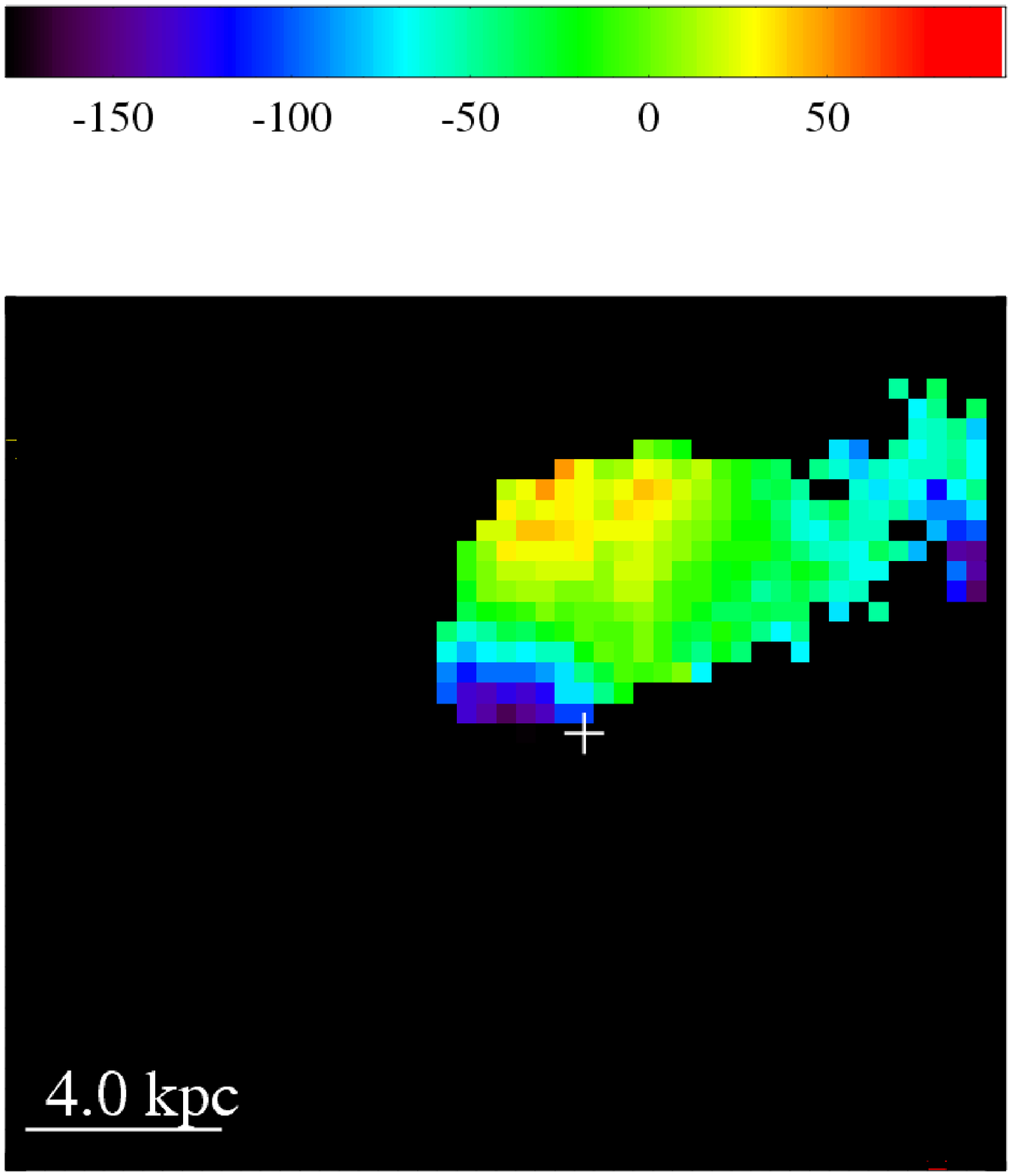}
  \includegraphics[width=0.4\columnwidth]{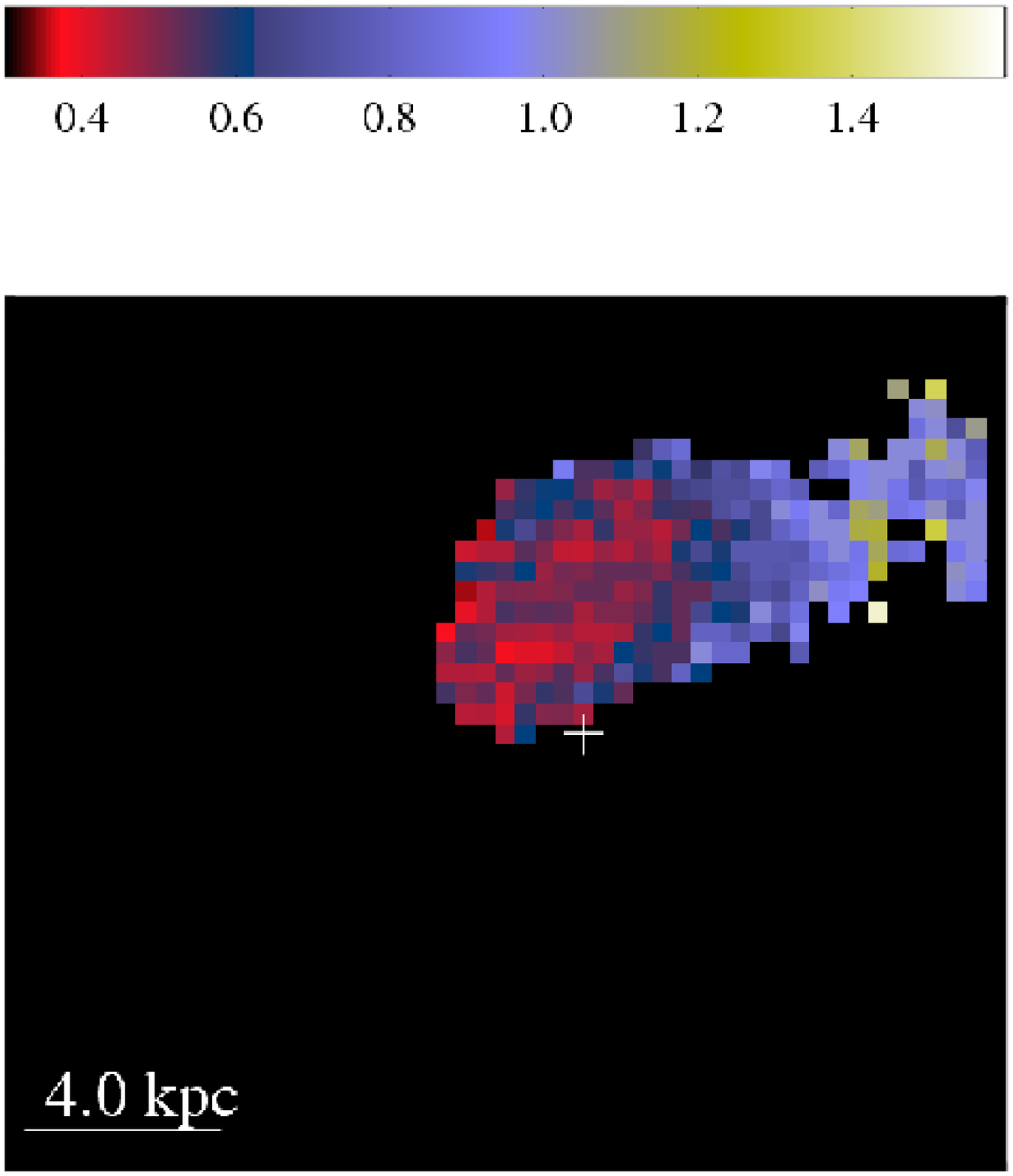}
   \caption{Images of RXJ\,0821 from left to right: HST WFPC2 F606w; normalized \ha +[N{\sc ii}] flux; line-width [\kmps]; line-of-sight velocity [\kmps]; [N{\sc ii}]/\ha\ line ratio. Images are 10.0$\times$8.4\,arcsec.  The cross marks the continuum peak which indicates the galaxy centre.
  \label{0821_ifu}}

    \includegraphics[width=0.452\columnwidth]{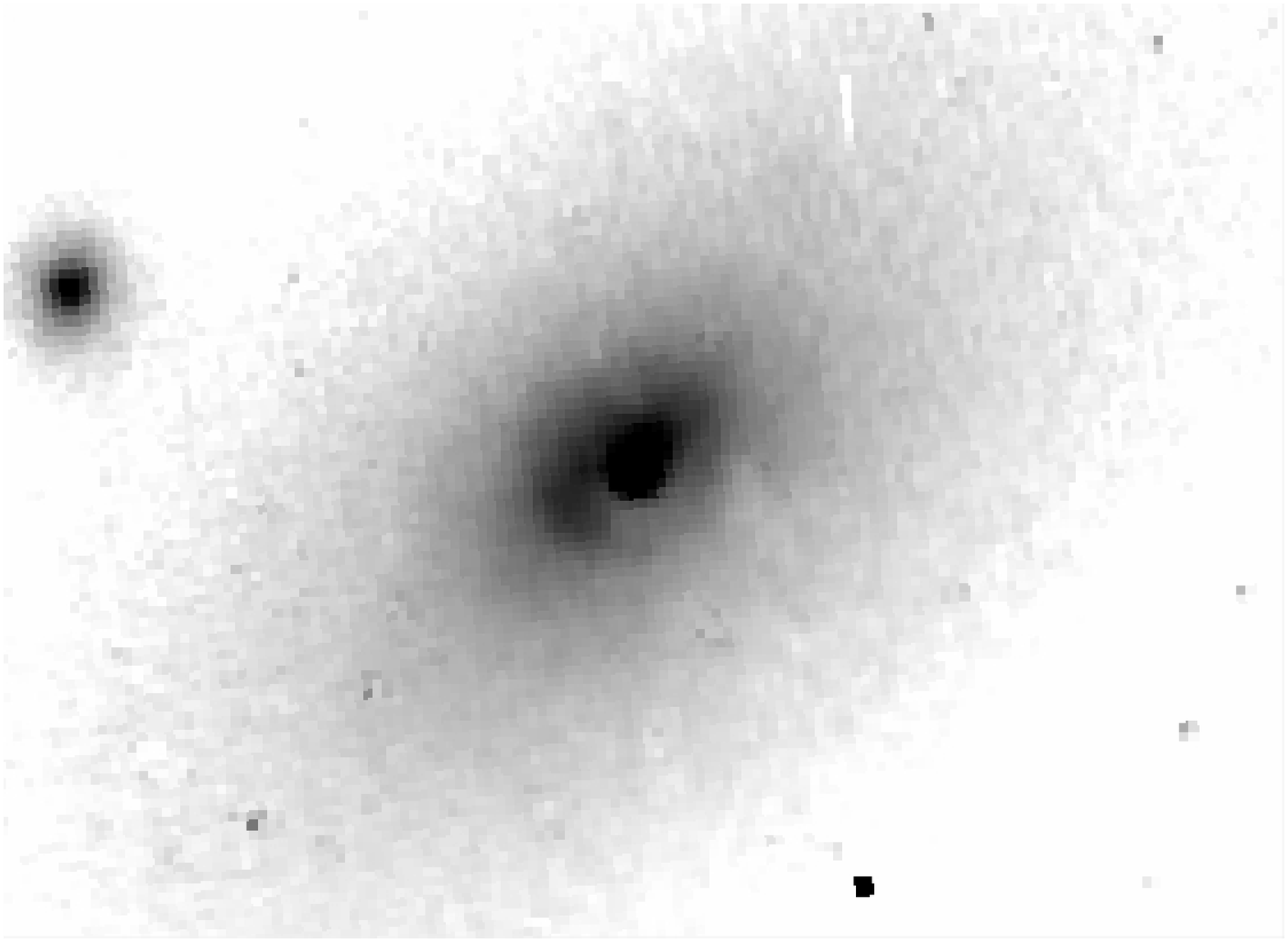}
    \includegraphics[width=0.4\columnwidth]{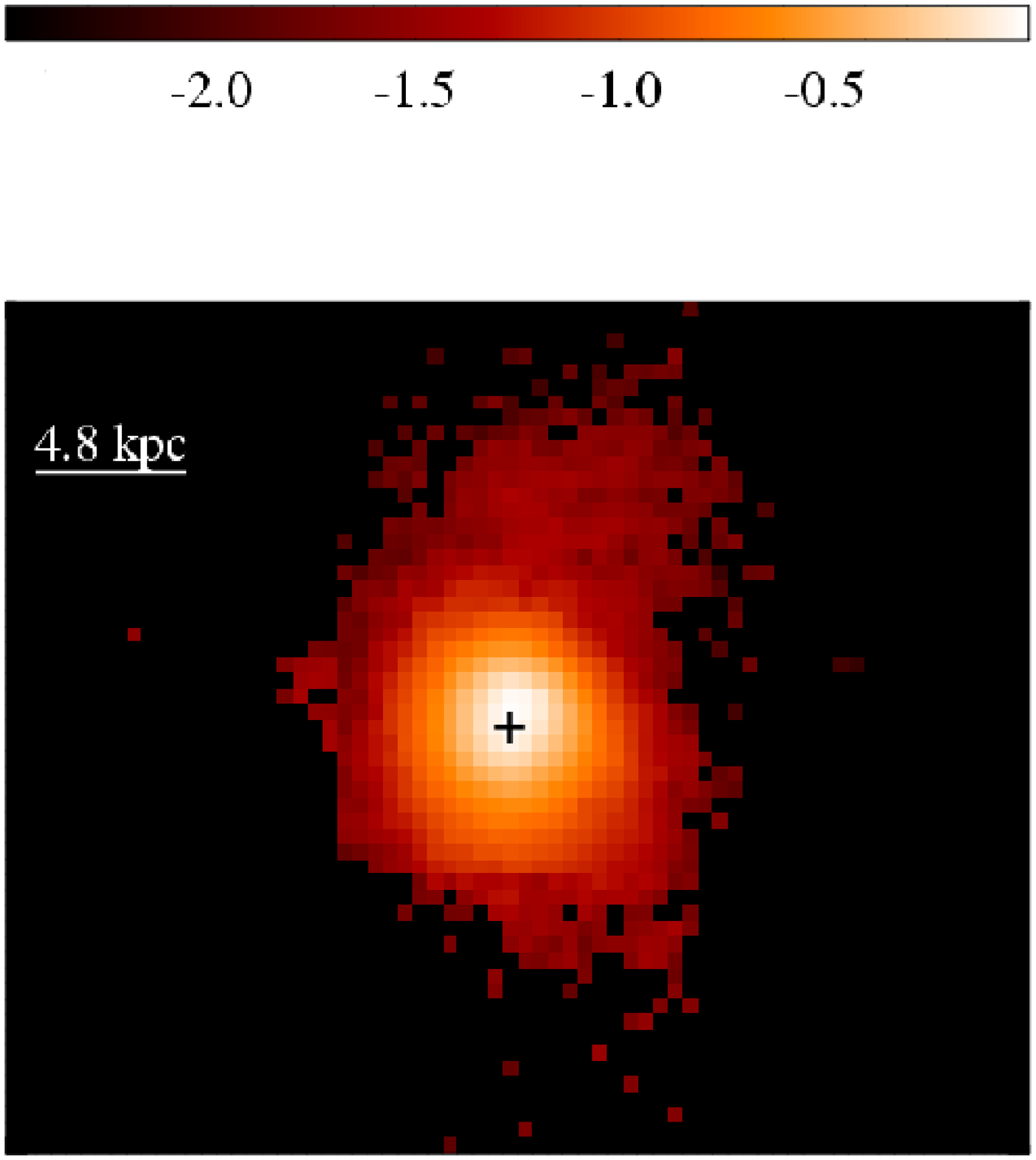}
  \includegraphics[width=0.4\columnwidth]{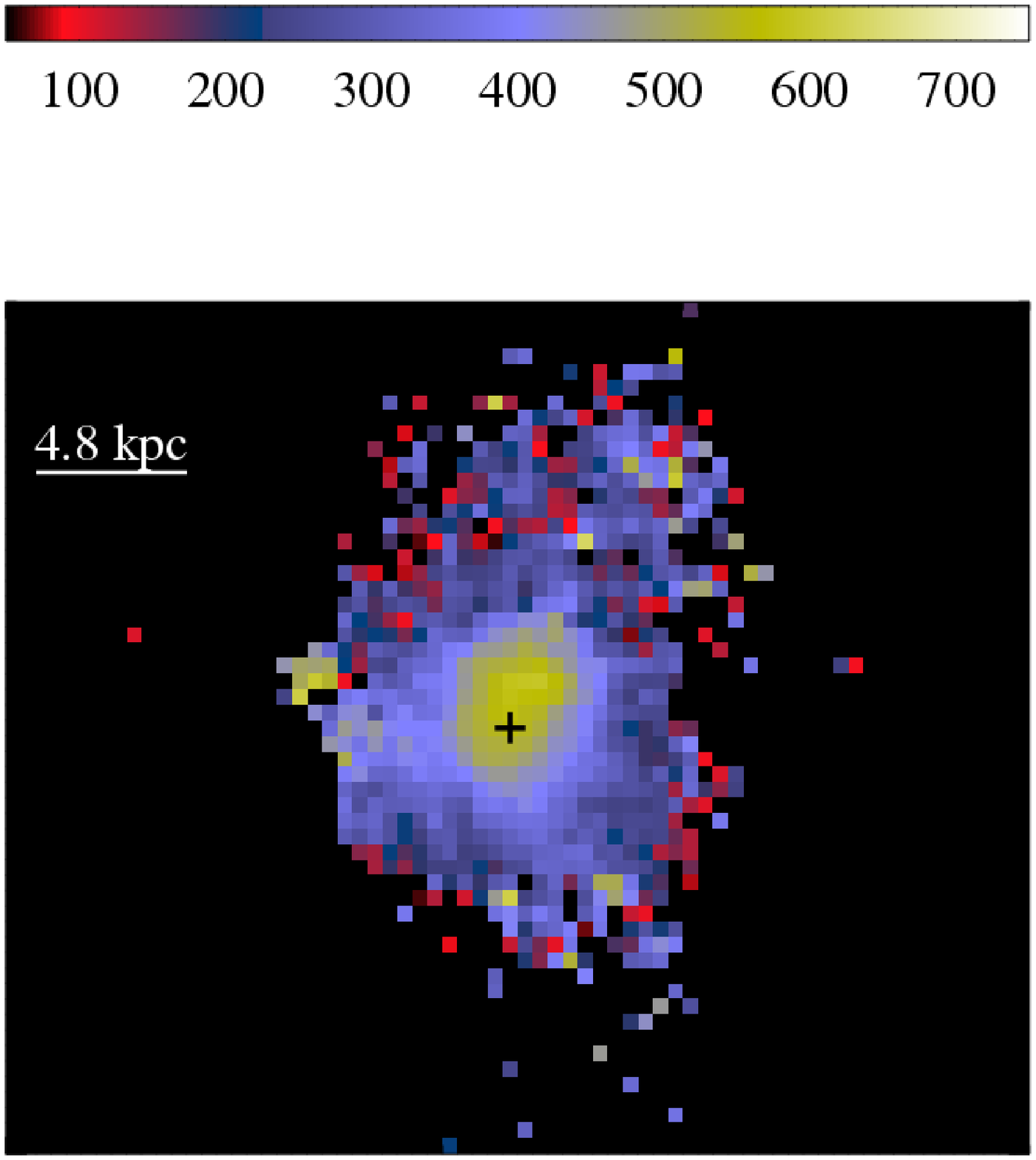}
\includegraphics[width=0.4\columnwidth]{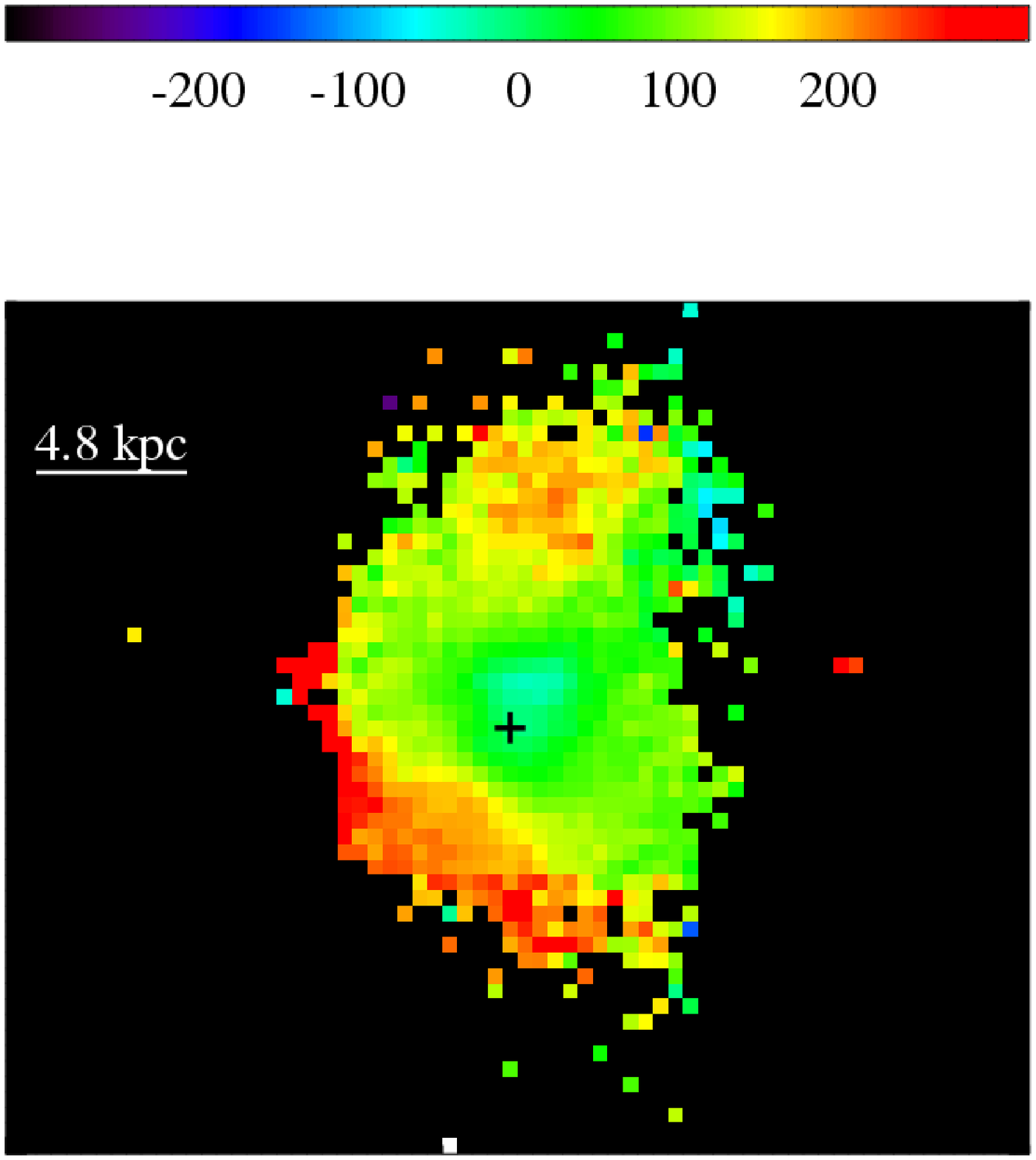}
\includegraphics[width=0.4\columnwidth]{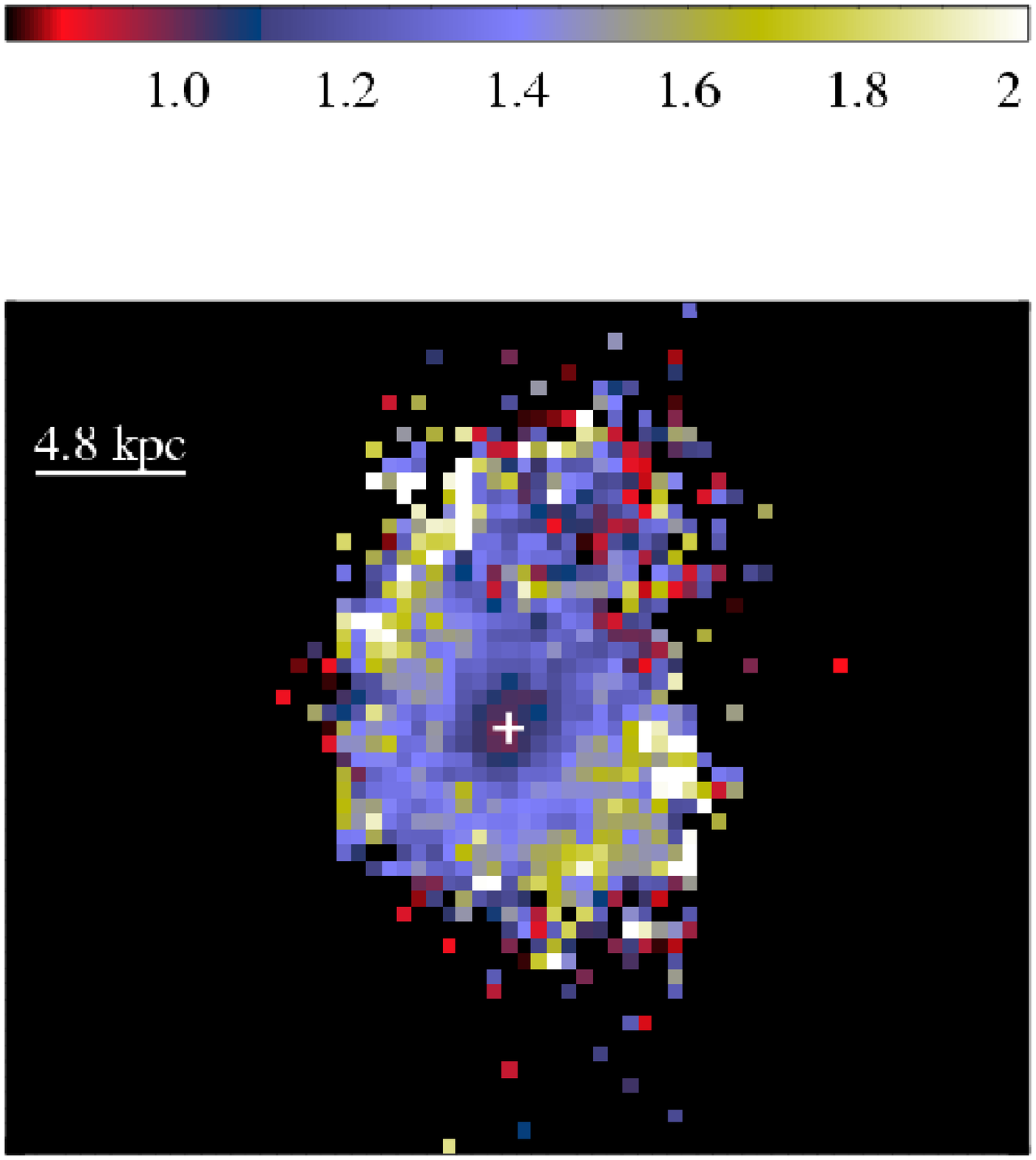}
   \caption{Images of A1068 from left to right: Negative HST ASC F606w; normalized \ha +[N{\sc ii}] flux (logarithmic scale); line-width [\kmps]; line-of-sight velocity [\kmps]; [N{\sc ii}]/\ha\ line ratio. Images are 13.4$\times$9.8\,arcsec.  The cross marks the continuum peak which indicates the galaxy centre.
   \label{A1068_ifu}}
\includegraphics[width=0.408\columnwidth]{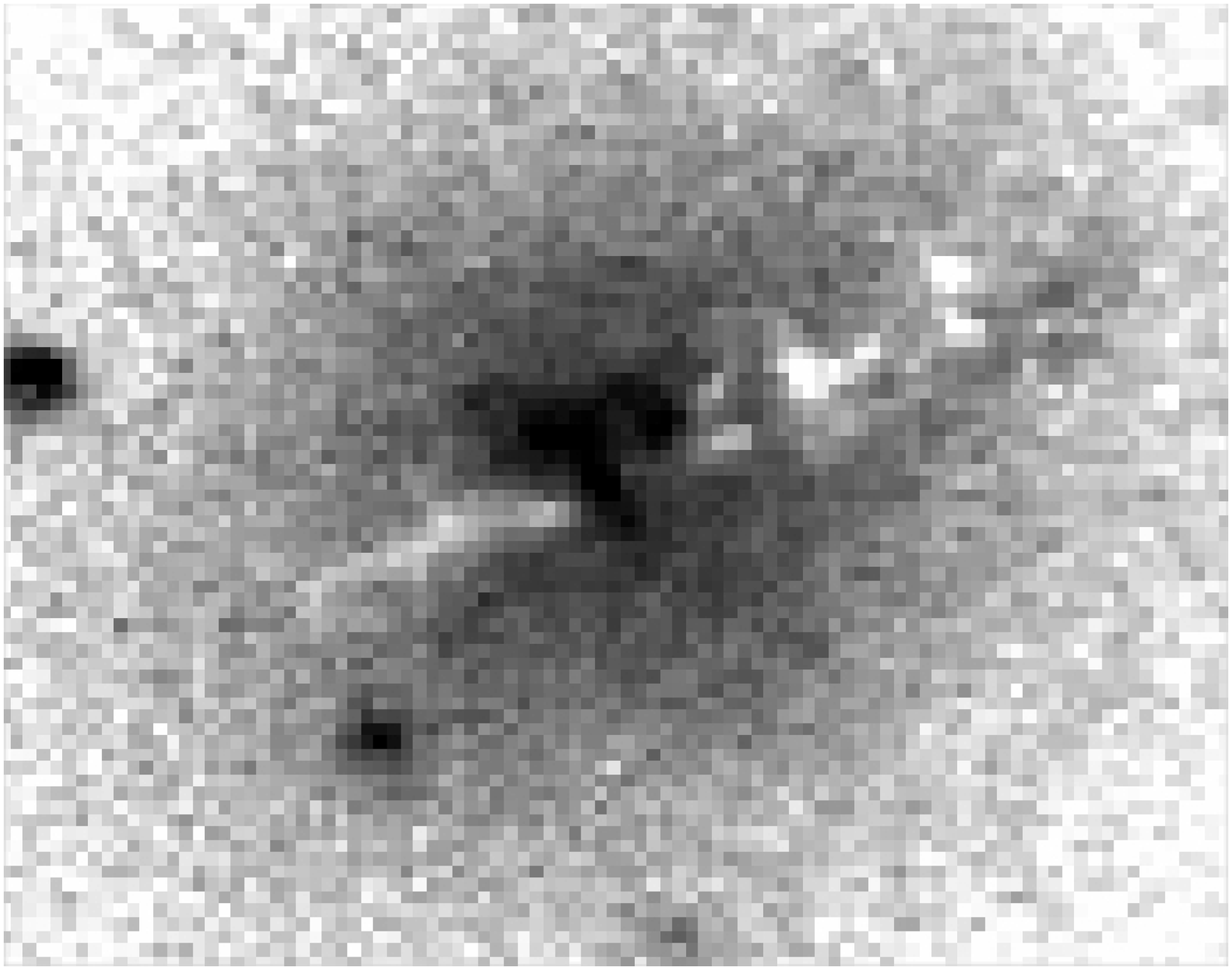}
\includegraphics[width=0.4\columnwidth]{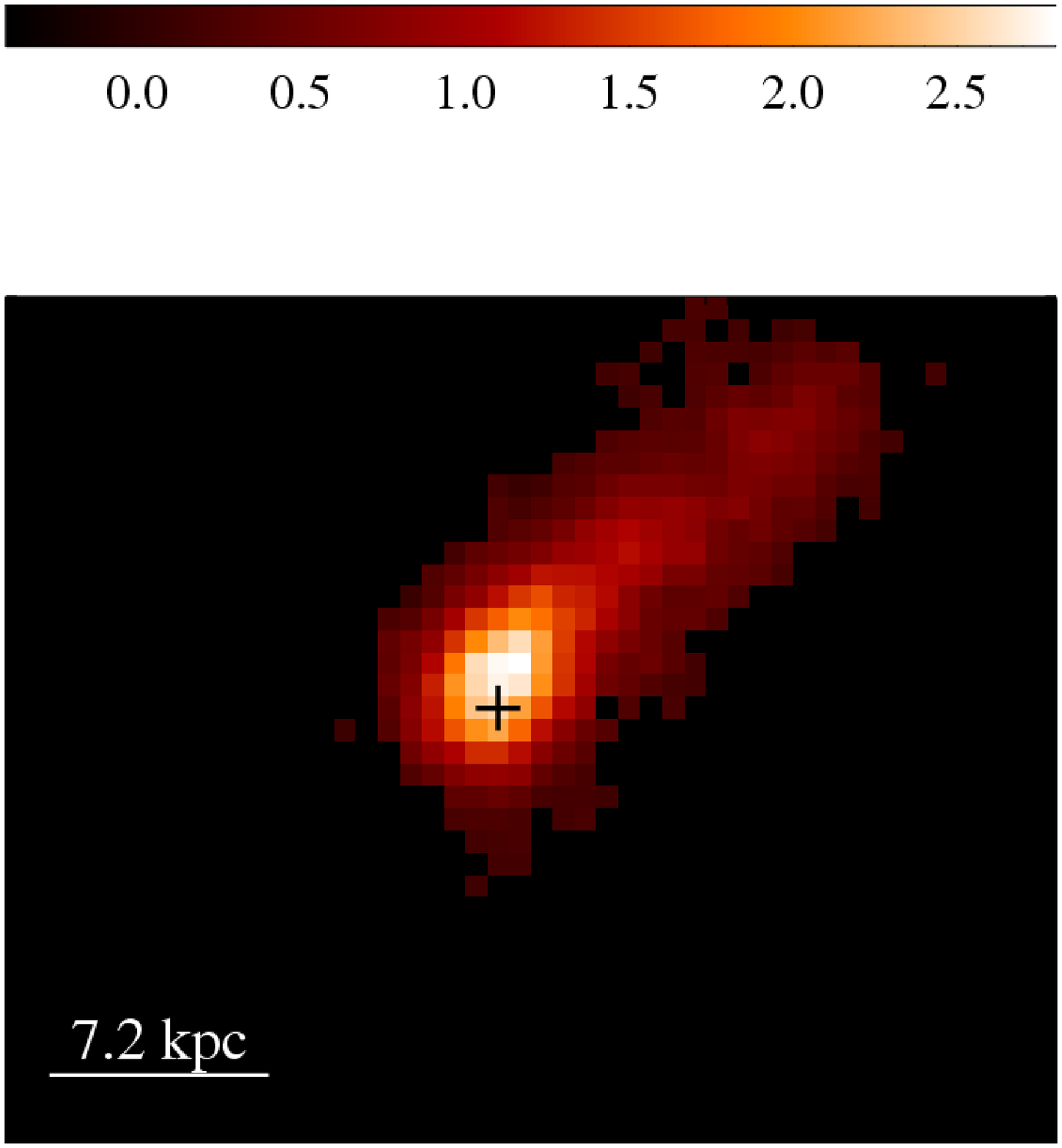}
\includegraphics[width=0.4\columnwidth]{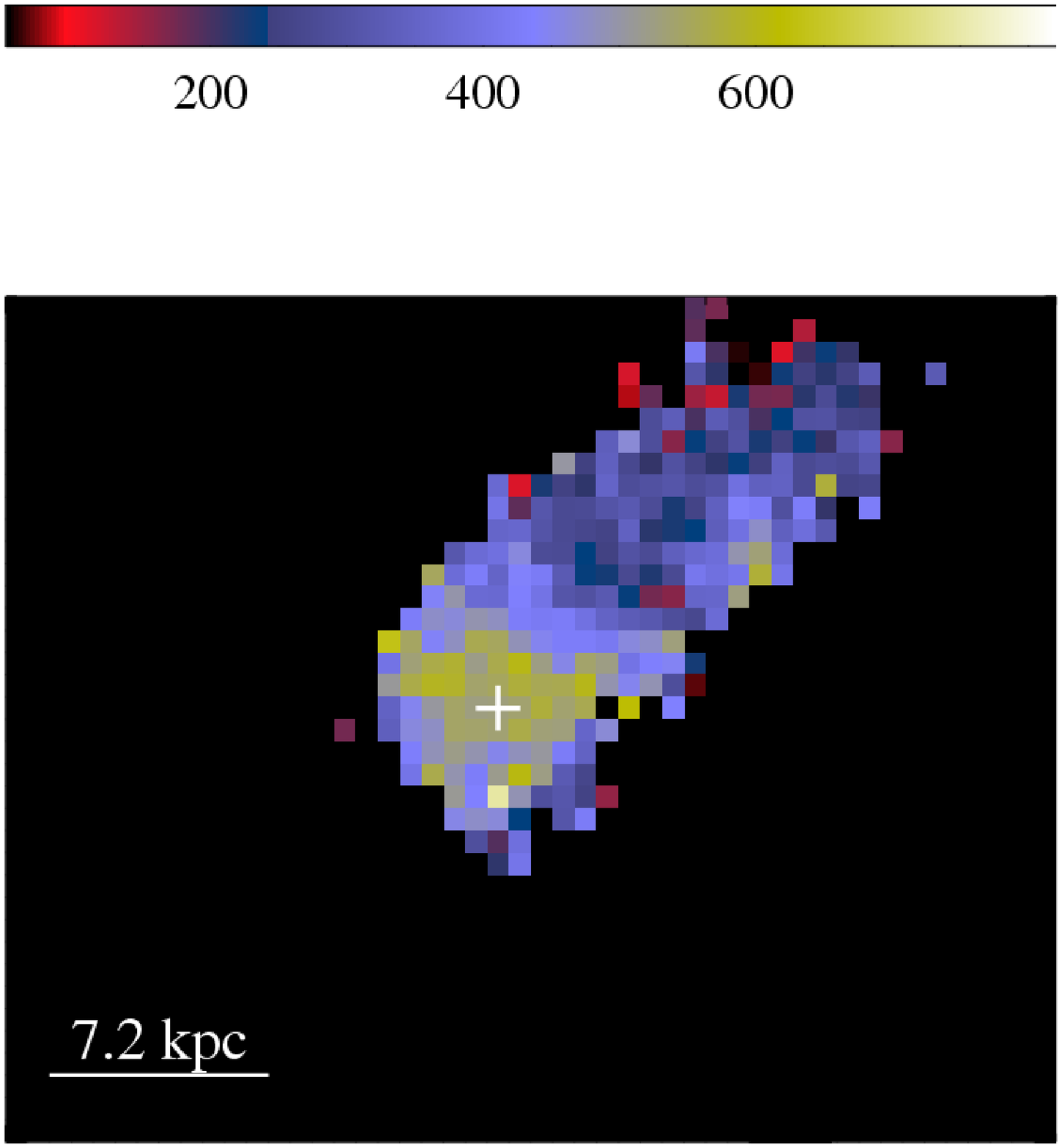}
\includegraphics[width=0.4\columnwidth]{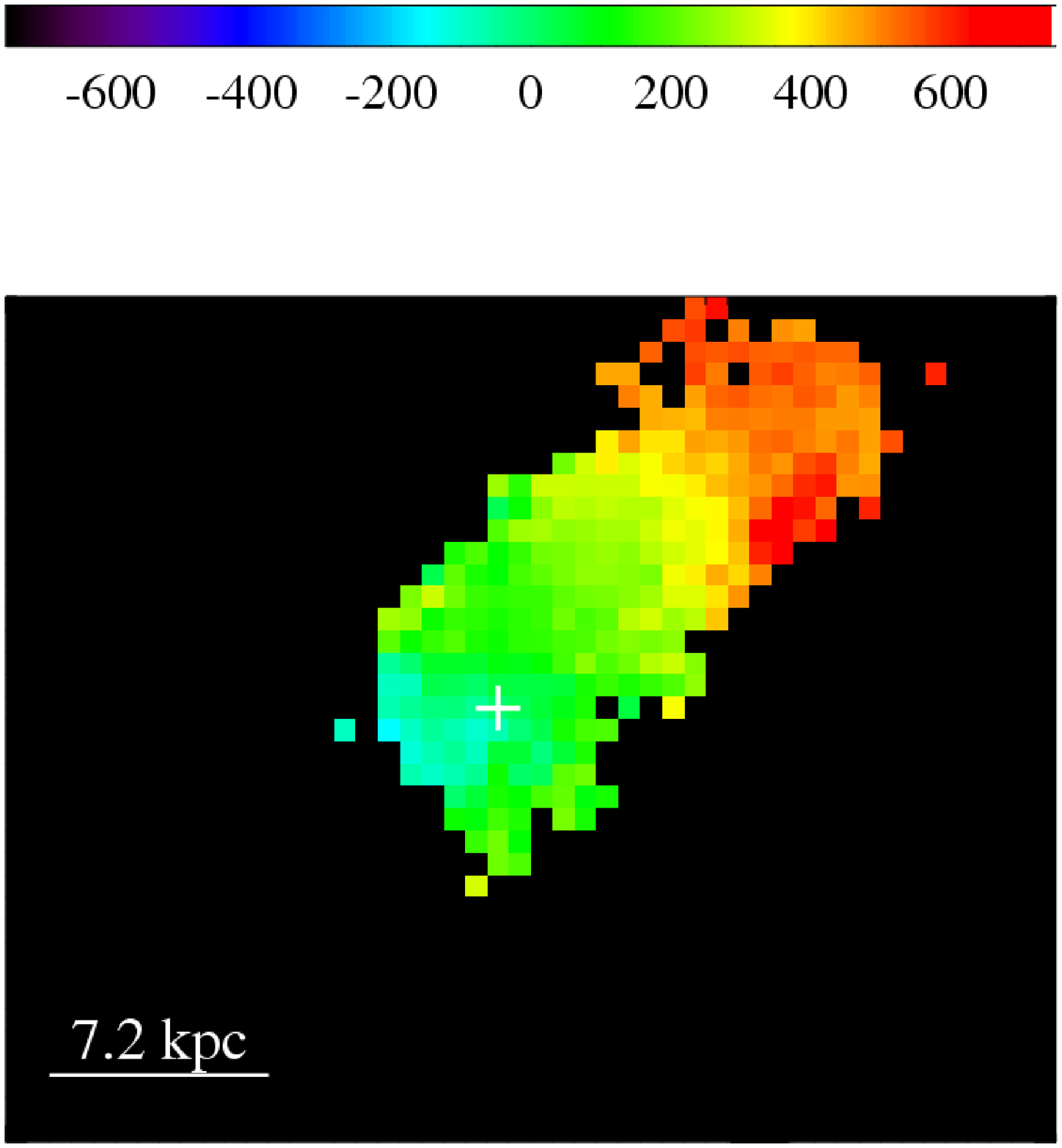}
\includegraphics[width=0.4\columnwidth]{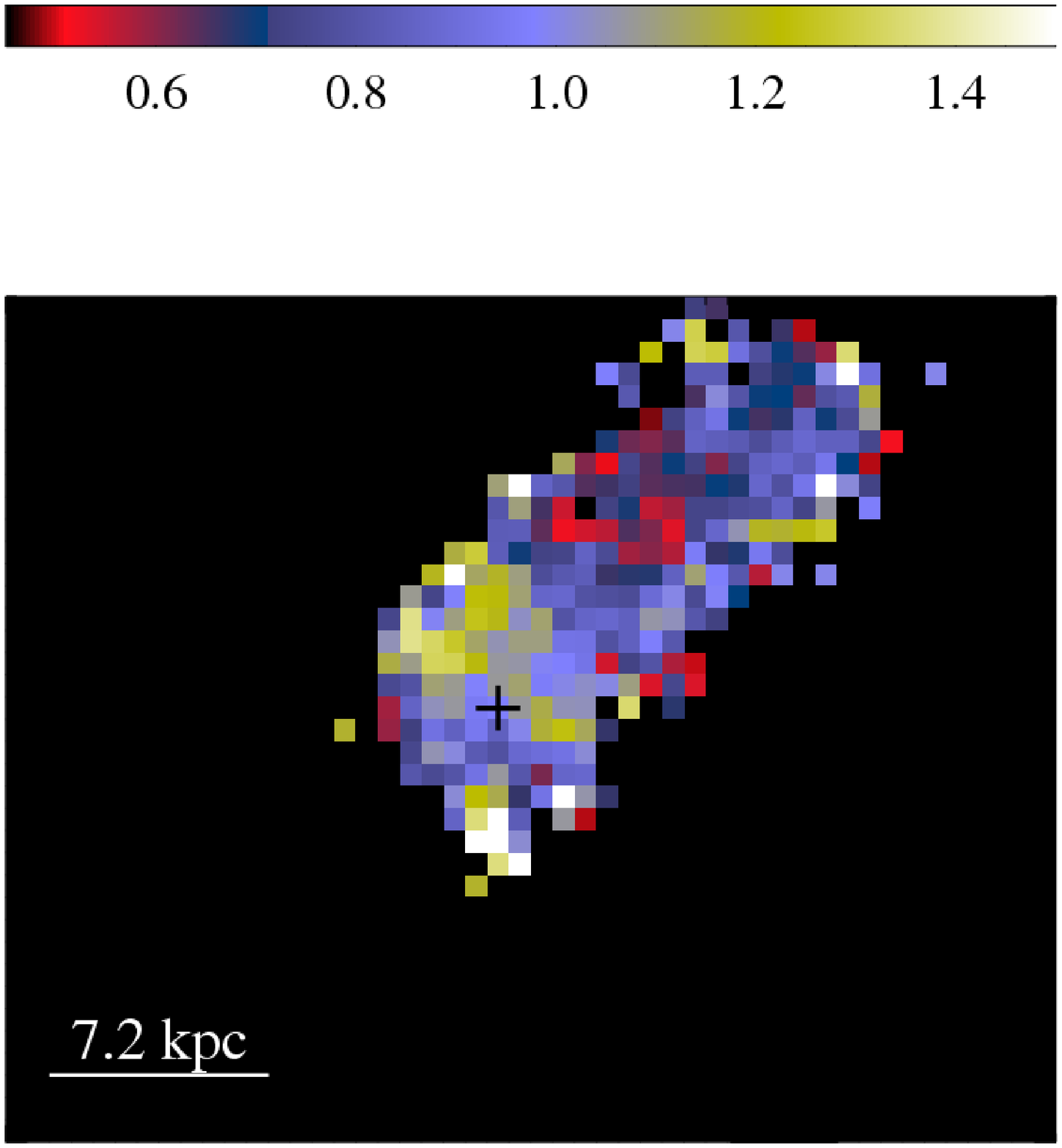}
\caption{Images of A2390 from left to right: HST ASC F555w/F814w; \ha +[N{\sc ii}] flux  (with colour bar scaling  in units of 10$^{-16}$\ergpspcmsq) ; line-width [\kmps]; line-of-sight velocity [\kmps]; [N{\sc ii}]/\ha\ line ratio. Images are 9.4$\times$7.4\,arcsec.  The cross marks the continuum peak which indicates the galaxy centre. \label{A2390_ifu}}
\end{figure*}

Fig.~\ref{A262_ifu} displays a Hubble Space Telescope (HST)
snapshot  and IFU images of the central galaxy of A262 including: \ha+[N{\sc ii}]
line-emission, line-width, line-of-sight velocity and [N{\sc ii}]/\ha\ ratio. The line-emission traces the large dust lanes that cut across
the galaxy including the main North-South dust lane and a secondary
component that stretches from the bottom left-hand corner to just beyond the galaxy
centre. There is further emission corresponding to dust patches in the
Northwest corner. At the centre of the nebula there is a bright
bar approximately 0.6\,kpc long at a position angle of
$122$\,degrees. The line-width is greatest at the centre of the galaxy but is elongated along the zero-velocity curve which is also the direction in which the radio emission emerges and orthogonal to the bright bar of \ha\ emission. Such kinematics may indicate unresolved rotation or an interaction between the radio jets and the ionized gas. The line-of-sight velocity of this region is low ($\le$100\kmps) and shows no bulk flow in the direction of the radio axis, but the gas motion is highly turbulent on scales of 0.2\,arcsec
(65pc). The ionized gas of A262 is rotating with a peak-to-peak velocity of 550\kmps.  $4\times10^{8}$\Msun\ of cold molecular hydrogen has been detected in this galaxy through the tracer CO \citep{Prandoni2006}.  The emission-lines of CO(1-0) and CO(2-1) have \lq double-horn' profiles with a FWHM of 550\kmps\ which indicates that the molecular gas exists in a rotating disk. The rotation velocity of the molecular gas is similar to the amplitude of the rotation seen in the ionized gas therefore the molecular and ionized gas are probably different temperature components of the same gas reservoir. Although the ionized gas is much easier to detect, it comprises only a tiny fraction (0.03\%) of the gas mass (see section \ref{denandpress}). An alternative interpretation is that the nebula is in a bipolar
outflow, but the nebula's association with the large dust lane argues
against this. If the nebula flows outward the receding half of the nebula should lie behind the bulk of the stellar light, therefore the clear association with dust would not be observed in the top section. 

The total \ha\ luminosity within the field-of-view of
the IFU is $L_{H_\alpha}=8.8\times10^{39}$ergs$^{-1}$. The [N{\sc ii}]/\ha\ line ratios
span the range of $1-4$. The brightest pixels, which lie in the central
region, have the lowest line ratios and the ratio tends to increase
radially, but the extreme outer regions to the Northwest and Southeast
also have very low ratios. 

\subsection{A496}
Fig.~\ref{A496_ifu} displays the IFU observations of A496 and a HST unsharped-mask snapshot of the galaxy. The unsharped-mask HST image shows 3
spiral dust lanes uncurling anticlockwise from the galaxy centre. The peak
\ha+[N{\sc ii}] emission corresponds to the area where the three dust
lanes meet in the galaxy centre. The line emission follows the general path
of the dust features, but is not as filamentary. Deeper high
resolution images are needed to determine whether the line emission is
directly associated with the dust, as in A262 or NGC\,4696 \citep{Crawford2005}.

The kinematics of A496 are smooth and ordered like A262. The
maximum line-width of $\sim$600\kmps occurs in the dust free central
region to the Northeast of the galaxy centre.  The rest of the nebula
has a line-width of 100$-250$\kmps.  The line-of-sight velocity
reveals a bulk flow, with the Southern part of the galaxy blueshifted
by $-$200\kmps whilst the Northern section is marginally redshifted up
to $+$150\kmps, but no clear kinematic pattern is associated with the
spiral dust structures.  Comparison of the emission-line kinematics
with the stellar kinematics presented in \citet{Fisher1995} show that
the two components are not connected. The peak-to peak gas velocity of
350\kmps is fairly low compared to the other BCGs, and the stellar
component of A496 has a mean rotation of only 29\kmps
\citep{Fisher1995}.  The low stellar kinematics are suggestive of a
lack of ordered motion within the stellar component, but the spiral
dust features prominent in the HST image and the bulk flow of the
nebula indicate that there is ordered motion in the gas component.

\subsection{2A~0335+096}

Fig~\ref{RXJ0335_ifu} displays the IFU observations and a HST snapshot  of
2A\,0335+096. Two separate \ha+[N{\sc ii}] central knots are visible
to the Northwest and Southeast of the nuclear continuum forming a
bar-like morphology that was first noted by \citet{Romanishin}. Bright
diffuse line-emission extends Northwest toward the secondary galaxy
that lies just beyond the IFU field-of-view, whilst dimmer
emission extends Northeast. The two \ha\ peaks of 2A\,0335+096 have different velocities. The Southeastern \ha\ peak is
blueshifted by $-$250\kmps compared to the Northwestern knot. The
zero-point of the line-of-sight velocity is uncertain due to the
presence of the two \ha\ peaks, neither of which match the peak in
continuum or line-width (although the continuum and line-width maxima are co-spatial). The two nuclei of 2A~0335+096 also have
different line ratios, the Southeast nucleus has larger [N{\sc ii}]/\ha\ ratios than the Northwest nucleus. Generally the [N{\sc ii}]/\ha\ ratio is large in the galaxy centre, decreasing with distance from the core.

The nebular gas that extends towards the secondary galaxy has a bulk velocity that is radially increasing in redshift.  The secondary galaxy is redshifted
by approximately 212\kmps relative to the central galaxy nucleus
\citep{Gelderman1996}.  Thus the Northwest emission not only extends
towards but also matches the velocity of the secondary galaxy.
\citet{Gelderman1996} also notes abrupt changes in line ratios, line
widths and radial velocity at the position of the secondary galaxy,
providing further evidence of an interaction. Given the
projected distance of The secondary galaxy (which lies 4.5\,kpc from the BCG nucleus)  may have disturbed the molecular gas reservoir approximately 30\,Myr ago (projected distance/ velocity shear) forming the large Northwest extension of the nebula.

\subsection{RXJ0821+0752}
Fig~\ref{0821_ifu} displays the IFU observations and a HST snapshot  of
RXJ\,0821+0752. Unlike the rest of the sample presented here the line emission is not primarily emitted from the galaxy nucleus (marked by a cross), but is offset Northwest of the galaxy. \citet{Bayer-Kim2002} have deeper spectra of the nucleus which show that there is very low surface brightness line emission that is below the detection limit of the IFU observations. The brightest part of the nebula is coincident with the bright arc and knots that surround the North of the galaxy (see the HST snapshot). These regions have strong blue continuum and stellar synthesis models have shown them to be star forming regions containing OB and A stars \citep{Bayer-Kim2002}.
There is a clear relationship between the \ha\ flux and the  [N{\sc ii}]/\ha\ ratio. Lenslets with \ha\ greater than half the maximum flux have [N{\sc ii}]/\ha $=0.5$. Below this \ha\ flux the ratio increases with 0.4$<$[N{\sc ii}]/\ha $<$1.2. Most low flux lenslets have a high [N{\sc ii}]/\ha. 

The largest line width of $\sim300$\kmps\ occurs in a region East of the nucleus that is coincident with low surface brightness diffuse continuum emission.  The arc and blue knots have a relatively low line width of 150\kmps\ and the line-width gradually decreases across the nebula from East to West.  As no line-emission was detected in the nucleus the line-of-sight zero point is defined as the lenslet with the peak \ha+[N{\sc ii}] flux. The line-of-sight velocity smoothly varies from $-180 - 90$\kmps. 
\subsection{A1068}
\label{A1068_distribution}

Fig.~\ref{A1068_ifu} presents a HST snapshot of A1068 together with the IFU observations. The IFU image captured
the majority of the emission-line nebula, but a faint Northwest
extension seen in the narrowband image of \citet{McNamara2004} is
below the IFU detection limit. The nebula extends Southwest almost
perpendicular to the stellar axis of the galaxy, and Northwest in the
direction of the radio emission. The core region
is extremely bright compared to the surrounding nebula, with a peak
surface brightness over 150 times greater than the majority of the
nebula. The central region shows signs of being in the
non-linear/saturated regime: the ratio of [N{\sc
ii}]$\lambda6584$/[N{\sc ii}]$\lambda6548$ within a radius
of 0.5\,arcsec around the central lenslet is less than 3 -- the value
set by the ratio of the statistical weights of the lines.  The brighter [N{\sc ii}]$\lambda6584$ line saturates before the dimmer [N{\sc ii}]$\lambda6548$, reducing the ratio from 3. The dimmer [N{\sc
ii}]$\lambda6548$ line is not saturated as the ratio only
decreases to 2 in the centre rather than 1. The \ha\ flux is slightly
less than the [N{\sc ii}]$\lambda$6584 flux but it is also likely to
suffer from nonlinearity effects and saturation in the core region. The decrease in the
[N{\sc ii}]/\ha\ ratio from $1.4$ to $1$ in the nucleus may be due to the saturation of the \nii\ line.  The core has also saturated in the HST image. A secondary galaxy lies to the Northeast, but the nebula does not extend towards it.

The line-width of the outer nebula is fairly uniform (at
$\sim100-200$\kmps) with a sharp increase towards the core, where the
line-width reaches 650\kmps. In this galaxy the maximum continuum, line-width
and line-flux peaks are co-spatial. There is a small
extension of the large line-width $\sim$20$^{\circ}$ West from North,
where the line emission is blueshifted by $\sim -50$\kmps relative to
the nucleus. This is the only region where blueshifted emission is
observed. Any outflow from the nucleus directed along the
line-of-sight would be masked out, as the zero-point of the
line-of-sight velocity is obtained from the lenslet at the centre of
the continuum peak.  In such a situation all the nebula would appear
redshifted compared to the nucleus. This appears to be the case in
A1068.  
\citet{McNamara2004} derive a star formation rate per unit area of
0.05--0.13\Msunpyrpkpcsq within a 10\,arcsec radius of the galaxy
centre. Starburst-driven winds are common in galaxies in which the
star-formation rate exceeds 0.1\Msunpyrpkpcsq \citep{Heckman2003}, therefore we may be viewing a starburst-driven outflow from the nucleus of A1068.  If the velocity zero-point is taken as an average of the whole nebula, then the outflow velocity
would not be greater than $\sim175$\kmps.  This value is relatively
low, especially when we consider that our viewing angle must be almost
directly aligned with the outflow. 

\citet{McNamara2004} observe UV flux  and infer significant star formation in the nucleus and the Northwest region. The  ionized gas in the Northwest and the nucleus  generally have lower [N{\sc ii}]/\ha\ ratios than at a comparable radius to the South. Therefore photoionization by massive stars plays a role in heating the gas and may lower the [N{\sc ii}]/\ha\ ratio.

\subsection{A2390}

Fig.~\ref{A2390_ifu} presents the IFU observations and a F555w/F814w HST image of A2390 which highlights the blue knots and dust features in this galaxy. Blue knots and dust features extend from the nucleus both Northwest and Southeast at a position
  angle of approximately $-45$ degrees. The nebula accompanies the blue light and dust to the Northwest; however, there is no emission-line gas to the Southeast. The peak of the \ha+[N{\sc ii}] line-emission coincides with  a dust lane that bisects the blue light morphology.  The black cross marking the peak of the continuum light corresponds to a bright blue  knot just below this dust feature. It is likely that the central dust lane is obscuring the stellar light causing the slight misalignment of the galaxy centre (defined by the continuum) and the peak of the line flux.  The total \ha\ luminosity from the galaxy is $L_{H_\alpha}=1.2\times10^{42}$\ergps.

The kinematics of A2390 are extremely ordered: the line-of-sight velocity gradually increases from the galaxy centre towards the Northwest.  The peak velocity of
$\sim$700\kmps is rather high compared to other BCGs (both in this
study and in \citealt{Heckman}).  The small extension to the Southeast
of the nucleus is blueshifted to $-100$\kmps.
\citet{Hutchings2000} measure a
Ly$\alpha$ velocity of $>$3000\kmps out to a distance of 1.5\,arcsec
from the nucleus which is much greater than the gas velocity measured from \ha\ in the
data presented here. These high-velocity components may contribute to
the unresolved kinematic component at the galaxy centre.

Such ordered motion my be due to normal galaxy rotation. Although it is unclear why we are able to observe only one side of this galaxy which has a peak-to-peak rotation amplitude of $\sim$1200\kmps. This rotation is hundreds of \kmps
greater than the rotation velocity typical of elliptical and S0 galaxies \citep{Sarzi2006}. Other possibilities include a cooling-wake; inflow; or outflow. 
A cooling-wake may be able to form when the BCG
oscillates around in the cluster potential. Gas may be stripped from
the galaxy and further gas cools directly from the ICM in a tail
behind the galaxy.  If the ionized gas within the nebula condenses
from the intracluster medium it should be kinematically linked to the
cluster rather than the central galaxy. Therefore the velocity of the
gas that is situated furthest from the galaxy should have settled to
the cluster velocity.  The nucleus of the central galaxy of A2390 has
a heliocentric radial velocity that is offset from the mean cluster
velocity (determined from 225 cluster galaxies; \citealp{struble1999})
by +590\kmps. The tip of the nebula is redshifted by almost +600\kmps
relative to the nucleus, which means it is offset by +1190\kmps
relative to the mean cluster velocity.  Thus the nebular kinematics suggests that it has not formed as a cooling-wake.

\citet{Hutchings2000} argue that it would be unusual to detect only
the receding (Northwest) side of a double jet system in \ha\ and  they attribute the blue knots and associated line-emission to infalling material. The line-of-sight speed increases with radius rather than decreases, as predicted by cooling flow theory \citep{FNC}, although this may be due to a projection effect if the filament is intrinsically curved along the lone-of-sight.

Alternatively the nebula may have formed from an outflow driven by a starburst or AGN. The galaxy is host to a relatively powerful radio source and the blue cones in the HST data have an opening angle of 20-30\deg, similar to other wide-angled AGN outflows
\citep{Veilleux2002}. However, the size and orientation of the
1.4\,GHz radio emission does not match the nebula.

 \citet{Crawford1999} derive a star formation rate of 0.18\Msunpyrpkpcsq (5.4\Msunpyr in an 2.3\,arcsec$^2$  aperture) by fitting stellar models to a long-slit optical
spectrum. Comparison to star formation measurements from other wavebands confirm this measurement: the LINER-like line-ratios mean it
is unlikely that all the \ha\ flux is produced by photoionization by
massive stars, but we can use the measured \ha\ luminosity as an upper
limit on the star formation rate.  The total \ha\ luminosity of
1.2$\times10^{42}$\ergps gives a star formation rate of 9.5\Msunpyr
\citep{Kennicutt1998}, and \citet{Egami2006} use {\it Spitzer} to
measure a star formation rate of 5\Msunpyr from the far-infrared
luminosity. A galaxy with a star formation rate $>$0.1\Msunpyrpkpcsq
is likely to produce a wind and therefore the nebula of A2390 may be
driven outward by the star formation.  The starburst model demands
that enough energy be produced by supernovae to drive the mass outflow
and supply the kinetic energy to the gas. To calculate the energy
released by the starburst we assume that 0.5\% of the stellar mass
formed will result in supernovae, and that each supernova event will
release 10$^{51}$\,erg, which will couple to the surrounding gas with
a 10\% efficiency \citep{Hutchings2000}.  Thus the star formation rate
of 5.4\Msunpyr will result in 0.027\,\Msunpyr\ of stars that will become supernovae. The mean supernova progenitor mass is at least $\sim$10\Msun, therefore the supernovae rate is 0.0027\,yr$^{-1}$ which results in an energy release into the surrounding gas of
$\sim3\times10^{47}$\ergpyr.  The total kinetic energy comprises both
the bulk and turbulent kinetic energy of the nebula, and amounts to
$E_{\rm kin}=1.45\times10^{55}$\,erg with the major component residing
in the turbulent kinetic energy ($E_{\rm
turbulent}=9.0\times10^{54}$\,erg) and the remainder in the bulk
kinetic energy ($E_{\rm bulk}=5.5\times10^{54}$\,erg). Therefore a star formation burst would have to last at least 50\,Myrs to provide the kinetic energy observed.
The energy required to work against the ICM pressure is insignificant
compared to the kinetic energy of the flow. The dynamical time of the wind ($\sim$15kpc/600\kmps) is $\sim$25\,Myr, and is therefore significantly shorter than the time required to accumulate the necessary energy. Therefore the nebular kinematics are consistent with an outflow powered by an AGN but not starburst-driven.

The range of line ratios in A2390 is much narrower than
the other galaxies in this sample, with the [N{\sc ii}]/\ha\ ratio
ranging between 0.5--1.4.  This
range matches well with NGC\,1275 and the high luminosity BCGs studied
by \citet{Wilman2006}. The dusty bar that bisects the galaxy at a
position angle of $\sim53$ degrees has the largest [N{\sc ii}]/\ha\ ratio of $1-1.4$.  The brightest blue
knot, approximately 2\,arcsec Northwest from the galaxy centre, has the lowest
[N{\sc ii}]/\ha\ ratio of $\sim0.6$. The rest of the extended nebula
has a [N{\sc ii}]/\ha\ ratio of $\sim0.8-1$.  As in A1068 and RXJ\,0821+0752, the
regions which show blue excesses have the lowest [N{\sc ii}]/\ha\ line ratio, whilst the
dusty regions have higher [N{\sc ii}]/\ha\ ratios. Different heating sources may be
operating in these separate regions producing the non-uniform line ratio. 
\subsection{Surface brightness profiles}
\label{sec:suface_brightness}
\begin{figure}
  \centering \includegraphics[width=1.0\columnwidth]{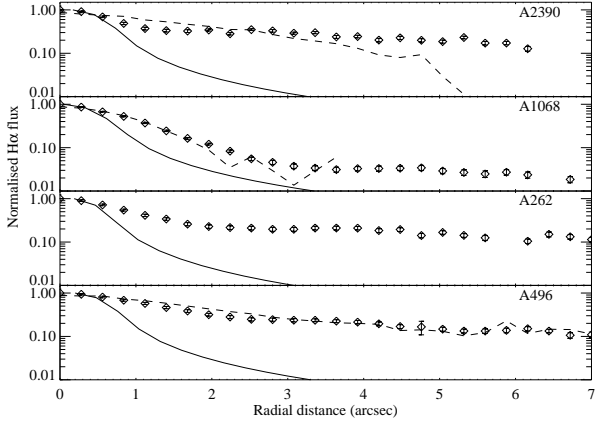}
  \caption{Normalised radially averaged \ha\ surface brightness profiles  (diamonds) of
  A2390, A1068, A262 and A496. The solid line is an inverse-square model convolved with a  Gaussian with a FWHM fixed by the seeing at time of observation, normalised to the second datapoint at 0.2\arcsec; the dashed line is the radial average of the continuum arbitrarily normailsed. Errors are plotted at the 1$\sigma$ level.}
\label{morph}
\end{figure}

The line-emission peaks close to the continuum peak except for 2A\,0335+096 and RXJ\,0821+0752. To check for the contributions from stellar-UV or AGN
ionization we compare the observed radial surface brightness
profile of \ha\ with two simple models. Ionization by the central AGN
will cause the radial \ha--flux profile to follow an inverse-square
law under the special conditions of constant gas
pressure and uniform cloud filling factor. If the \ha\ clouds are in
hydrostatic equilibrium with the surrounding ICM then
$\frac{\rm dP}{\rm dr}<0$ and the \ha\ flux should drop with a steeper
gradient than inverse-square. It the filling factor of the clouds is
non-uniform, the inverse-squared model is invalid. Ionization by stellar light will cause the radial flux profile to
be more complex, but in its simplest form it is expected to follow the
continuum profile. The dust absorption in A262 makes a comparison
between the line-emission and the stellar light difficult, therefore
the line-flux profile is not compared to a stellar ionizing model. The
radial profile of both the line flux and continuum is created by
averaging the \ha\ flux and continuum in circular apertures centred on
the central lenslet (marked by a cross).  Fig.~\ref{morph} plots the
radial \ha\ profiles of the BCGs within clusters A262, A496, A1068 and
A2390. The \ha\ surface brightness profile for all four galaxies is significantly broader than the inverse-square model, therefore it is unlikely that the nebulae are ionized
by a central source such as an AGN beyond a radius of 0.5\,arcsec. The \ha\ profile follows the continuum profile in A496 and A1068 therefore photoionization by stellar UV may be important in these extended nebulae.

\subsection{Density, pressure and mass}
\label{denandpress}
The nebulae of A262, A496, A2390 and A1068 exhibit strong [S{\sc ii}]$\lambda$6717, 6731 emission allowing an estimate to be placed on the electron density.
The density is measured from the average of  lenslets where both the [S{\sc ii}] lines are detected above 3$\sigma$.   For nebulae where the electron density exceeds the lower limit probed by the [S{\sc ii}] doublet (100\,cm$^{-3}$) we estimate the pressure assuming that the region containing the S$^{+}$ is 50\% ionized so that the total gas density is three times the electron density, and that the gas temperature is
10,000\,K. Table \ref{density-pressure} lists the average densities and
pressures. Although there are large errors associated with the ionized gas
pressures, they are very similar to the X-ray-derived ICM pressure.
\citet{Heckman} found that the ionized gas was overpressurized by an
order of magnitude compared to the ICM, although the authors stress
that the X-ray-derived gas pressures were unreliable.  The
measurements of the ICM and ionized gas pressure are now converging,
partly due to the spatial resolution offered by Chandra which has
enabled more accurate measurements to be made of the ICM thermal
pressure close to the nebula.
\begin{table}{
\centering
\begin{tabular}{lllll} \hline
&A262&A496&A1068&A2390 \\ \hline 
Density  & 400 $\pm^{300}_{150}$&100 ($<2000)$&650$\pm^{1350}_{400}$&600$\pm^{800}_{350}$\\
Nebular pressure &1.7$\times10^{-9}$&4.1$\times10^{-10}$& 2.7$\times10^{-9}$&2.5$\times10^{-9}$\\
 ICM Pressure &1$\times10^{-10}$&4$\times10^{-10}$&6.5$\times10^{-9}$&1.7$\times10^{-9}$\\
\ha\ luminosity&8.8$\times 10^{39}$&--&--&1.2$\times 10^{42}$\\
Mass (\Msun)&1.2$\times 10^5$&--&--&5.6$\times 10^6$\\ \hline
\end{tabular}
\caption{\label{density-pressure} Average density  (cm$^{-3}$), pressure (dynes\,cm$^{-2}$) and \ha\ luminosity  (\ergps) estimates of the nebulae for which the [S{\sc ii}] doublet is
  observed. Central ICM pressure from  \citet{Blanton2004}, \citet{Wise2004},
  \citet{Allen2001} and \citet{Dupke2003}}}
\end{table}
\label{sec:mass_A2390}

The masses of the ionized gas from A2390 and A262 are obtained from the \ha\ luminosity through
\begin{equation}
{\rm M}= \frac{L({\rm H}\alpha)m_{p}}{n_{e}\alpha^{eff}_{{\rm H}\alpha}h\nu_{H\alpha}}~,
\end{equation}
where $n_{e}$ is the number density of electrons, $\alpha^{eff}_{{\rm
H}\alpha}$ is the effective recombination coefficient for \ha\ line
emission and $h\nu_{\rm H\alpha}$ is the energy from a photon at the
frequency of \ha\ \citep{Osterbrock}. We assume case B recombination and
a nebular temperature of 10,000\,K to derive the masses listed in Table \ref{density-pressure}.

Assuming pressure equilibrium with the surrounding ICM we can estimate a filling factor for the ionized gas. Densities vary between $100-1000$\,cm$^{-3}$, therefore the observed  $1-10 \times10^{5}$\Msun\ of ionized gas ($\sim10^{62}-10^{63}$ atoms) must occupy $10^{60}$\,cm$^{3}$ which corresponds to a sphere of radius 30\,pc. The emission we observe is spread over a large fraction of the IFU field-of-view and occupying at least 3kpc$^{-3}$, therefore the volume filling factor must be  1$\times10^{-9}$ or less. It is likely that the nebulae exist in the form of a web of filaments such as those seen in nearby objects (e.g.~Perseus; \citealt{Conselice})

\subsection{[N{\sc ii}]/\ha\ ratio}
\begin{figure}
\centering
\includegraphics[width=1\columnwidth]{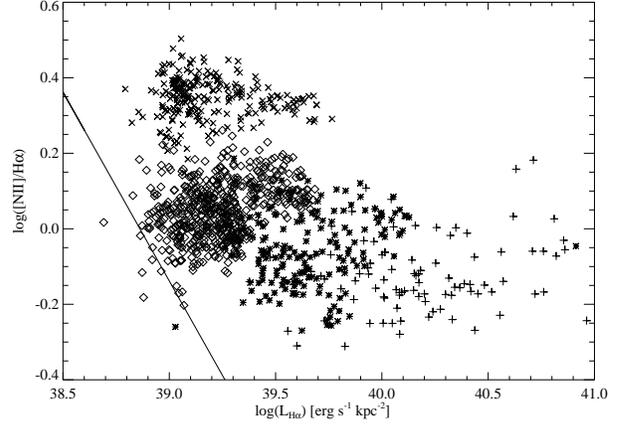}
\caption{\label{n2-ha} \nii/\ha\ verses \ha\ luminosity per square kpc
  for A262 (X), A2390 (asterisks), 2A~0335+096 (diamonds) and
  NGC\,1275 (crosses).  Datapoints are restricted to those in which
  \ha\ and \nii\ are detected above 9$\sigma$. The solid line marks
  the 9$\sigma$ detection limit.}
\end{figure}
The forbidden lines such as \nii\ result from the excitation of N$^{+}$ through collisions with thermalized electrons that were liberated through photoionization.  The \ha\ emission results from the recombination of the hydrogen ion.  Whilst the rate of photoionization depends on the strength of the radiation, the mean energy of the liberated electrons (which determines the nebular temperature) only depends on the form of the ionizing radiation field. The total collisional rate of the electrons with N$^{+}$ ions depends on the nebular temperature as well as the electron and ion densities, therefore the \nii\ flux depends on the N$^{+}$ abundance, the strength of the radiation field and the form of the radiation field: a harder ionizing source will produce a greater \nii\ flux. In ionization equilibrium the number of photoionizations is balanced by the number of recombinations, thus the \ha\ flux depends on the strength of the radiation field. The ratio of the forbidden \nii\ to \ha\ line will depend on the metallicity of the gas and the form of the ionizing radiation.

There is a correlation between the \nii/\ha\ ratio and the total nebular \ha\ luminosity of BCGs \citep{Crawford1999} where high luminosity nebulae have low [N{\sc ii}]/\ha\ ratios. Within individual nebulae some trends between the line ratio and \ha\ luminosity are clear in the IFU  [N{\sc ii}]/\ha\ images. A1068 and  RXJ\,0821+0752 both show a decrease in  [N{\sc ii}]/\ha\ with \ha\ flux, whilst A2390 shows the opposite trend of increasing  [N{\sc ii}]/\ha\ with \ha\ flux. To test whether there is any general trend between \ha\ flux and [N{\sc ii}]/\ha\ in individual regions within the nebulae we plot  \nii/\ha\ verses \ha\ surface brightness for all flux calibrated datasets (A262, 2A~0335+096, A2390 and NGC\,1275 [data obtained from \citealt{Hatch2006}]) in Fig.~\ref{n2-ha}.  The solid line marks the average 9$\sigma$ detection limit.  No trend exists between the [N{\sc ii}]/\ha\ ratio and \ha\ surface brightness.  Instead,
datapoints from highly-luminous nebulae tend to lie towards the bottom
of the graph with a small range in line ratio around [N{\sc
ii}]/\ha$\approx$1, whilst A262 (a relatively low-luminosity nebula)
has a larger spread of line ratios which tend to be higher. The form of the ionizing radiation and/or the gas metallicity are not uniform but must vary within each galaxy and between the whole sample.

\section{Discussion}
\label{discussion}
\subsection{Dusty nebulae}
The low \ha-luminosity nebulae A262 and A496, are co-spatial with absorbing
dust lanes clearly visible in the HST snapshots.  A2390, 2A~0335+096, RXJ\,0821+0752, and A1068 are more luminous \ha\ nebulae, situated in clusters with large X-ray luminosities and have
comparatively large star formation rates.  These nebulae have a larger
spatial extent and do not correspond with dust structures, although
many nebulae without visible dust structures are known to be dusty \citep{Donahue1993,Edge1999,Egami2006}. Dust does not have a long survival time in the hot ICM \citep{Draine1979}. It can only form when the gas is shielded from strong UV and X-rays, and lies near regions rich in stars.   BCGs are known to contain vast
reservoirs of molecular hydrogen with column densities of
$10^{22}$\,cm$^{-2}$ \citep{Edge2001,Edge2003,Salome2003} which can shield the gas from the ICM X-rays long enough for it to
become polluted with dust. Therefore the dusty nebulae have probably been drawn out of the molecular reservoirs that lie in the core of BCGs. The dust in the nebulae will be destroyed by sputtering by X-rays from the surrounding ICM on a timescale of $\sim10^{6}(a_{\micron}/n_{ICM})$\,yrs \citep{Draine1979}, where $a_{\micron}$ is the grain size in microns and $n_{ICM})$ is the density of the surrounding ICM in cm$^{-3}$ ($\sim0.1$cm$^{-3}$). The kinematics suggest that the nebulae are at least $5\times10^7$ \,yrs old (see section \ref{kin}), however they still contain dust. If the dust has been destroyed by sputtering, only the big dust grains ($>5\micron$) will remain.  \citet{Sparks1989} showed that the properties of the large dust lane in the emission-line nebula of the BCG NGC\,4696 were similar to dust in the Milky Way, therefore the dust grains have a similar grain-size distribution. This rules out the possibility that the sputtering has removed the dust grains smaller than 5$\micron$ in this BCG. Therefore the dust must be shielded from the X-rays and possibly exists in small dense clumps. Further studies of the properties of dust in BCGs are needed to confirm whether the grain-size distribution is similar to the Milky Way, or whether the smaller dust grains have been destroyed. 
\subsection{Nebular kinematics}
\label{kin}
The kinematics of the nebulae suggest that tidal interactions between nearby galaxies and the central molecular reservoir, AGN- and starburst-driven outflows may all
play a role in their formation. The central galaxy of 2A~0335+096 appears to have been disturbed by the nearby
secondary galaxy. The surrounding nebula extends towards the secondary galaxy that
lies 6.5\,arcsec away in projection and increases in velocity smoothly from the BCG nucleus to the secondary galaxy. Four other BCGs have galaxies that appear nearby in projection (A1068, A496, RXJ\,0821+0752 and A2390), but their nebulae do not extend towards these galaxies nor do their kinematics indicate that interactions have recently taken place.  A2390 and A1068 show signs of outflow, possibly starburst or
AGN-driven. 

A262 and A496 have kinematics consistent with rotation. The A262 nebula rotates with the same rotation speed as the underlying molecular gas reservoir. Therefore these nebulae are merely the ionized skins of the molecular reservoir in these galaxies and do not require any disturbance of the reservoir to become ionized. 
Bulk subsonic motions and shear of a few hundred \kmps across lengths of a few kpc are seen in the more luminous nebulae. They do not have ordered rotation and are not in stable configurations. Therefore these nebulae may disperse, or assuming the molecular reservoir provides fuel for the nebulae,  change morphology on the order of the dynamical timescale (approximately 5$\times10^7$yrs). The smooth velocity gradients of a few hundred \kmps\ across distances of $\sim$10\,kpc imply that the ionized filaments have formed over $\sim$50\,Myrs. Thus the ionized nebulae are long-lived and must be able to survive evaporation by the hot ICM for at least this time. 

The line width maps of the ionized gas are fairly uniform with a central gradient near the nucleus of the galaxy with FWHM of 600-800\kmps, comparable to the velocity dispersion of the central regions of elliptical and lenticular galaxies \citep{Sarzi2006}. The ionized gas can be pressure supported in these central regions if the large line widths are the result of turbulent motions. The line width of the gas beyond this central region is typically  $50-150$\kmps, which is much greater than the expected thermal broadening of gas at 10,000\,K  ($\sim$10\kmps). Infrared studies have shown that the ionized gas is  accompanied by large amounts of molecular hydrogen at all radii \citep{Hatch,Jaffe2005,Johnstone2007}: the gas is denser and more massive than observations of the ionized gas phase indicate. Therefore the line width is too low to provide pressure support against the gravitational potential of the $\sim10^{12}$\Msun\ galaxy. The extended regions of the ionized nebulae in A262 and A496 can be supported by rotation, however the other galaxies in this sample lack organised rotation.  Long-lived stable gas needs to be supported, without rotational or pressure support the nebula should collapse on timescales of $\sim10^7$yr, reaching free-fall velocities up to $\sim$2500\kmps.  The radial velocities are almost an order of magnitude below the free-fall velocity, therefore the nebulae must be supported and we must appeal to magnetic pressure support or the gas may be dragged by the moving ICM as seen in NGC\,1275 \citep{Fabian2003,Hatch2006}.  Alternatively the ionized gas, which must have been drawn out from the galaxy centre (since it contains dust and molecular hydrogen; \citealt{Hatch,Johnstone2007}), may disperse into the ICM before it falls back to the galaxy. As it does so, the nebula can pollute the ICM with metals from the galaxy centre at large radial distances. 
\subsection{Source of nebular heating}
The heating source of the atomic gas has been examined in almost every
study of these nebulae, but it remains uncertain. The large [N{\sc
ii}]/\ha\ and low [O{\sc iii}]/\ha\ ratios have proved problematic: no
photoionization or shock model can reproduce the observed
spectra. Within each of the six nebulae we found significant spatial
variation of [N{\sc ii}]/\ha. Most pronounced is the [N{\sc
ii}]/\ha\ map of A2390 that shows direct correspondence between the
[N{\sc ii}]/\ha\ line ratio, star forming blue knots (lowest [N{\sc
ii}]/\ha), and a dust lane (highest [N{\sc ii}]/\ha). In
A1068 the [N{\sc ii}]/\ha\ ratio is lower in regions which have a large UV
flux than in the remainder of the nebula, and the bright knots in RXJ\,0821+0752 that have blue continuum also have the lowest  [N{\sc ii}]/\ha\ ratio and highest \ha\ flux. Even though the spectral features of the nebulae do not match photoionization models, the
decrease in the [N{\sc ii}]/\ha\ ratio in regions of UV or blue light
excess implies that stellar UV enhances the \ha\ emission affecting
the line ratio. UV from massive stars is a contributing heating
mechanism in certain areas, but the data also imply that this source
of heating must accompany an underlying heating source that produces
the high forbidden-to-recombination line ratios.

The spatial variation of the line ratios implies
the ionization state of the gas is not uniform.  Different excitation
mechanisms may act in different regions.  This is in contrast to the
sample of distant (z$>0.13$), and highly luminous nebulae observed by
\citet{Wilman2006}, where the authors observe a generally uniform
ionization state.  \citet{Wilman2006} suggest that the optical line
emission from high luminosity nebulae is predominantly powered by
distributed star-formation, induced by a perturbation of the central
BCG gas reservoir, which produces uniform line ratios across the
nebula.  They predict that lower luminosity systems, which do not have
such intense star-formation, will show more variation in the [N{\sc
ii}]/\ha\ ratio, with the ratio increasing away from the star-forming
clumps.  This trend is indeed observed in the high luminosity nebulae RXJ\,0821+0752 and A2390.  If star-formation is assumed to be the only mechanism boosting the \ha\
flux and decreasing the [N{\sc ii}]/\ha\ ratio, we would expect to see
a trend between the [N{\sc ii}]/\ha\ ratio and the local \ha\ surface
brightness, similar to the correlation between the [N{\sc ii}]/\ha\
ratio and the total \ha\ luminosity \citep{Crawford1999}.  No such
trend is found in the complete flux calibrated sample nor in the sample of
\citet{Wilman2006}. A slight correlation between the \nii/\ha\ ratio and \ha\ flux for the non-flux calibrated nebulae of RXJ\,0821+0752 and A1068, but notably not in A2390. Therefore there are likely to be other factors
affecting the line ratios; for example, large variations in the gas
metallicity, or multiple excitation mechanisms which may act with
varying degrees in different nebulae.

The optical and near-infrared emission-lines point towards a hard and
distributed ionizing source.  Extremely hot ($>10^5$\,K)
Wolf-Rayet stars can produce the correct line ratios \citep{Jaffe2005}. However, the Wolf-Rayet phase of the massive star lifetime lasts only $0.3-0.7$\,Myr
\citep{Schaller1992}, so it is unlikely that Wolf-Rayet stars can
provide continuous heating throughout the extended nebula over the
lifetime of the filaments, which the nebular kinematics suggest is at
least tens of millions of years.  Wolf-Rayet galaxies, which contain a
large population of Wolf-Rayet stars, tend to have broad He\,{\sc
ii}$\lambda$4686 emission features. Although these features are
sometimes found in the centre of BCGs with high star formation rates
(e.g.~A1835; \citealt{Allen1992}), the He\,{\sc ii}$\lambda$4686
emission feature is generally not observed in the majority of these
nebulae.  There is a correlation between \ha\ emission and star formation rate in BCGs \citep{Johnstone1987,Allen95}, therefore it is tempting to assume that the star formation can ionize the nebulae. This only works if the majority of the UV radiation from the massive stars is absorbed by the nebula, which will occur only if the nebula has a covering fraction of unity. The observations presented here do not rule this out, however this scenario cannot be true for nearby nebula for which we have high resolution images that show a covering fraction $<<1$.

Stellar UV is not the only plausible locally ionizing source, the surrounding ICM is another possible source of energy and other suggested mechanisms include turbulent mixing layers \citep{Crawfordmixinglayers}, and heat conduction \citep{Sparks1989} or ionization by the thermal electrons from the ICM. The mean free path ($\lambda=10^{4}T^{2}/n$ in cm; \citealt{Cowie1977}) of a thermal electron from the hot ICM (10$^7$\,K) in the nebular gas (n$_e\sim$100\,cm$^{-3}$) is $\sim$0.3pc. Therefore heating by the thermal electrons from the ICM is a plausible mechanism providing that the clumps of  \ha\ emitting gas are greater than this size. A further important unknown is the magnetic field geometry which could prevent particles from the hotter phase penetrating to the much cooler one. Magnetic fields are required within the clouds to prevent them being shredded by the hot gas (see \citealt{Loewenstein1990} for a discussion of magnetic field strengths for various cloud sizes). The nearby nebulae in M87, NGC\,1275 and the Centaurus cluster show soft X-ray emission from the same location as the line emitting filamentary nebulae \citep{Sparks,Fabian2003,Crawford2005} which suggests an energy exchange between the ICM and the nebular gas. The resulting spectrum from such a heating source has been examined by \citet{Bohringer1989} for a steady state model who found that the resulting flux in the line emission fell below typical nebular values.  

 The kinematics rule out uninterrupted free-fall of the nebula to the galaxy centre, therefore the ionized gas must be slowed down, possibly by magnetic field lines, and in doing so Alfven heating \citep{Loewenstein1990} may play a role in heating the nebular gas. Spitzer observations of two nearby nebulae have shown that cool molecular hydrogen ($\sim$330\,K) coexist with the ionized gas at all radii, and $\sim20$ times more mass exists in the molecular phase than the ionized phase \citep{Johnstone2007}.  Therefore the total mass in the nebulae may be in excess of $10^{8}$\Msun, leading to a great deal of potential energy being released as the nebular gas falls back to the galaxy.

\section{Conclusions}
We have presented IFU observations mapping the morphology, kinematics
and ionization state of six nebulae that surround BCGs at the centre
of clusters with cool cores. The low-luminosity nebulae A262 and A496
are co-spatial with large dust lanes, but the high-luminosity nebulae
extend beyond the visible dust features. Smooth velocity shears and
bulk flows of a few hundred \kmps\ occur in all nebulae with 
no general trend in the kinematics. Instead multiple phenomena may affect the nebular motion, including interactions with nearby galaxies, as seen in 2A~0335+096, AGN or starburst driven outflows as postulated in A2390 and A1068 respectively. One nebula (A262) shows no signs of disturbance as it has the same rotation velocity as the underlying molecular reservoir. Therefore the existance of a nebula may not rely on a disturbance of the molecular reservoir, but merely excitation of the gas. 

The nebulae have low-ionization spectra, with large [N{\sc ii}]/\ha\ line ratios, but the nebular ionization state is not uniform, regions with excess blue or UV light have lower [N{\sc ii}]/\ha\ line ratios. Therefore stellar UV influences the line ratio but is not the dominant ionization source in many nebulae; there must exist another distributed,  continuous source of heating that produces the low-ionization spectra. 
\section{Acknowledgments} 
We wish to thank the anonymous referee for useful comments and a careful reading of the manuscript. CSC and ACF thank the Royal Society for financial support. The data were obtained on WHT which is operated on the island of La Palma by the Isaac Newton Group in the Spanish Observatorio del Roque de los Muchachos of the Instituto de Astrof'sica de Canarias. We wish to thank the WHT team, in particular Sam Rix and Ian Skillen for taking the bulk of the observations. 
\bibliographystyle{mn2e}\bibliography{mn-jour,CCG}

\end{document}